\documentclass[%
 reprint,
 amsmath,amssymb,
 aps,
floatfix,
]{revtex4-2}

\usepackage{setspace}
\usepackage{hyperref}
\usepackage{graphicx}
\usepackage{dcolumn}
\usepackage{bm}
\usepackage[export]{adjustbox}

\usepackage[
margin=1in,
]{geometry}
\usepackage{siunitx}
\usepackage{csquotes}
\usepackage{array}
\usepackage{url}

\begin{document}

\title{Determination of non-local characteristics of density transport in 2D simulations of the SOL}

\author{T. Gheorghiu$^{1, 2}$} 
\email{theo.gheorghiu@ukaea.uk}
\author{F. Militello$^{1}$}
\author{J. Juul Rasmussen$^{3}$}
\affiliation{$^1$United Kingdom Atomic Energy Authority, Culham Campus, Abingdon, OX14 3DB, UK \\
$^2$York Plasma Institute, Department of Physics, University of York, Heslington, York YO105DD, UK \\
$^3$Physics Department, Technical University of Denmark, DK-2800 Kgs. Lyngby, Denmark}

\date{\today}

\begin{abstract}
By use of Lagrangian tracers propagated on 2D simulations of Scrape-Off Layer (SOL) turbulence, we are able to determine the non-local fractional-advection, fractional-diffusion equation (FADE) coefficients for a number of equilibrium cases. Solutions of the resultant FADEs shows good agreement with the simulated mean density profiles. We detail how the FADE is derived: the stochastic flux equation is introduced, and it is shown how it is used to find general forms of Fick's first and second laws, dependent on the the jump function. We show for spatially homogeneous jump functions which belong to the Levy-$\alpha$ Stable distribution - that transport may be approximated by a non-local FADE with four parameters. This work demonstrates the sound basis for FADEs to act as reduced models of transport in systems dominated by coherent structures; so justifies the development of a first-principles approach to calculating FADE parameters.
\end{abstract}

\maketitle

\newpage

\section{\label{sec:1} Introduction} 
Understanding density transport in the scrape-off layer (SOL) of tokamak plasmas is critical for predicting and optimizing plasma confinement and performance in fusion devices. Transport in the SOL is influenced by the turbulent formation, ejection, and radial propagation of coherent structures known as filaments or blobs \cite{Zweben1985, Zweben2004}. We then seek to develop reduced models of radial transport that can take into account the phenomenology of the SOL. 

The initial approaches adopted to characterise radial SOL transport are effective advection-diffusion equations \cite{Garcia2007b} - these traditional transport models implicitly assume local behaviour based on Fick's laws, which implies linear flux-gradient relations. However, it has become clear that turbulence and the presence of coherent structures in the SOL leads to cases where there is no clear flux-gradient relation \cite{Garcia2007, Garcia2007b, Naulin2007}, and more recently attempts have been made to model the SOL using a non-local turbulence spreading approach \cite{Yan2021, Long2024} - correspondingly, these cases cannot be described with a simple advection-diffusion relation, and so we must develop a different way of characterising radial transport behaviour. 

There has been recent interest in using stochastic approaches \cite{Garcia2012, Losada2022, Militello2016b} to model the SOL - while this approach has seen some success, it is reliant on assumptions about the statistical properties of filaments. 

In previous work \cite{Gheorghiu2024}, we developed an observational random walk approach; this approach assumes the existence of a jump function and corresponding observation interval, which when Gaussian leads to the classic advection-diffusion equation - but if non-Gaussian, would indicate that different model equations would be required to describe transport. By measuring jump functions of Lagrangian tracers in turbulence modelled by the Hasegawa-Wakatani equations, we were able to demonstrate that jump functions were non-Gaussian in systems dominated by coherent structures - indicating non-locality.

Previous work on non-local transport in fusion plasmas has been conducted, with experimental work indicating radial transport of heat as non-local \cite{Gentle1995}, and attempts to describe heat flux using a non-local kernel method \cite{Pradalier2010}. Apparent non-locality in the core is reviewed by Ida \cite{Ida2022}. Fractional models of transport have been proposed before \cite{CastilloNegrete2006, CastilloNegrete2008}, however they lacked a clear first principles justification for their use.

In this work, we present a general non-local model of transport that allows us to characterise a system with a set of four effective parameters. This approach allows us to retain the appeal of the advection-diffusion approach - the ability to characterise systems with effective transport parameters - while still being able to model the radial transport of density in the SOL. This is done through the construction of a stochastic flux equation, which is then used to derive a fractional-advection, fractional-diffusion equation (FADE), which can model non-local transport. We then use jump functions obtained from the motion of Lagrangian tracers on 2D simulations of midplane SOL turbulence using the STORM module of BOUT++ \cite{BOUTv4-3-2, Nicholas2021} to identify the FADE parameters, and finally compare the solutions with the simulated SOL density profiles to assess their suitability as a reduced transport model. 

In section \ref{sec:2} we construct a generalized Fick's law using a stochastic flux approach. In section \ref{sec:3} we derive the fractional-advection, fractional-diffusion equation (FADE) using the generalized Fick's law and the generalised central limit theorem (GCLT). In section \ref{sec:4} we discuss the STORM2D equations used to simulate the midplane SOL, as well as the method of obtaining jump functions from the simulation. In section \ref{sec:5}, we compare the simulated mean profiles to the solutions of the FADEs obtained from the measured jump functions.

\section{\label{sec:2} Stochastic Flux}
It is useful to construct a general Fick's law that can help us understand and model non-Fickian fluxes. We state Fick's first law:

\begin{equation}
    \label{eq:2.1}
    \Gamma(x) = - D \frac{\partial P(x)}{\partial x}
\end{equation}

Here $\Gamma(x)$ is the flux at a point x, $D$ is the diffusion coefficient, and $P(x)$ is a background field such as density. The general form of Fick's first law will be defined in relation to the background field and the jump function, $q(\Delta x)$. The jump function is defined as the probability of a particle having a measured spatial displacement $\Delta x$, if the time between two observations of the particle is $\tau$. The limitations on the jump function for the purposes of this paper are that it should be spatially homogeneous and invariant with time. It is possible to extend the analysis to a spatially inhomogeneous and time-dependent jump function - however, without understanding how to construct the jump function from physical properties of the system, it is not clear how we should permit the jump function to vary in space and time in a way that reflects reality.

Given particles emanate from a point a distance $x_n$ from a surface $w$ in a stochastic manner dictated by the jump function, then over a time $\tau$ the flux over that surface due to that point may be written as: 

\begin{equation}
    \label{eq:2.2}
       \Gamma_{w, L}(x, x_n, t) \equiv \frac{1}{\tau} P(x - x_n, t) \int^{+\infty}_{x_n} q(\Delta x) d \Delta x
\end{equation}

Where it is assumed that $\tau$ is sufficiently long such that any fluctuations such as those due to turbulence may be neglected. $\Gamma_{w, L}$ represents the ``left" contribution to the flux. The ``right" contribution due to a similarly placed point on the other side of the surface is:

\begin{equation}
    \label{eq:2.3}
    \Gamma_{w, R}(x, x_n, t) \equiv \frac{1}{\tau} P(x + x_n, t) \int^{+\infty}_{x_n} q(-\Delta x) d \Delta x
\end{equation}

The net instantaneous flux across the surface due to the pair of points $x \pm x_n$ is then:

\begin{equation}
    \label{eq:2.4}
    \Gamma_{w}(x, x_n, t) = \Gamma_{w, L}(x, x_n, t) - \Gamma_{w, R}(x, x_n, t)
\end{equation}

Therefore the total flux across the surface is that due to all pairs of points: 

\begin{equation}
    \label{eq:2.6}
    \tau \Gamma_{w}(x, t) = \int^{+\infty}_0 \Gamma_{w, L}(x, x_n, t) - \Gamma_{w, R}(x, x_n, t)~dx_n
\end{equation}

Eqn.~\ref{eq:2.6} is the instantaneous stochastic flux, as it explicitly relates the flux at a point to the jump function and the background field; this can then be used to obtain Eqn.~\ref{eq:2.1}, which occurs for the case that the jump function is Gaussian. 

The Fourier transform pair of Eqn.~\ref{eq:2.6} may be found \cite{Gheorghiu2024b} to be: 

\begin{equation}
    \label{eq:2.8}
    \tau \hat{\Gamma}(k, t) =  \hat{P}(k,t)  \frac{i}{k} \left[ \hat{q}(k) - \hat{q}(0) \right]
\end{equation}

The Fourier transform is defined as in Eqn.~\ref{eq:2.7} with corresponding inverse. 

\begin{equation}
    \label{eq:2.7}
    FT\left\{f(x)\right\} = \hat{f}(k) = \int^{\infty}_{-\infty} f(x)e^{-ikx} dx
\end{equation}

Eqn.~\ref{eq:2.8} is then a general Fick's first law, relating the flux to both the jump function and background field. Applying the divergence theorem in the absence of sources or sinks:

\begin{equation}
    \label{eq:2.9}
    \frac{\partial P}{\partial t} + \nabla \cdot \Gamma = 0
\end{equation}

...gives us a general Fick's second law: 

\begin{equation}
    \label{eq:2.10}
    \tau \frac{\partial \hat{P}}{\partial t} = \hat{P}(k,t)   \left[ \hat{q}(k) - \hat{q}(0) \right]
\end{equation}

Should the jump function be a probability density function (PDF), then its Fourier pair should have the property $\hat{q}(0) = 1$ \cite{Nolan2020}. 

Note that the Eqn.~\ref{eq:2.6} allows direct relation between the jump function, and kernels used in other attempts at characterizing non-local transport \cite{Pradalier2010}.

\section{\label{sec:3} The Fractional-Advection, Fractional-Diffusion Equation}
The choice of $\hat{q}(k)$ determines the form of the general Fick's laws via Eqn.~\ref{eq:2.6}. The general central limit theorem (GCLT) \cite{Gnedenko1952} implies the existence of the stable, or Levy-$\alpha$ stable (L$\alpha$S) \cite{Nolan2020} distribution as the limit of convergence of a particular series of random variables. Where the central limit theorem (CLT) deals with the distributional convergence of a normalized sequence of independent, identically distributed (iid) random variables with finite variance, the GCLT deals with the distributional convergence given a sequence of iid random variables only - as such, it is of broader applicability. 

In the case of density transport, given equilibrium, we expect the jump function of particle displacements to have converged to some time invariant distribution. Consequently, it is reasonable to expect the jump function to belong to the L$\alpha$S distribution. The L$\alpha$S distribution is given in Fourier space as in Eqn.~\ref{eq:3.1} \cite{Nolan2020}: 

\begin{widetext}
\begin{equation}
    \label{eq:3.1}
    L\alpha S(k; \alpha, \gamma, \beta, \delta) = 
    \begin{cases}
        e^{-ik\delta}e^{-|\gamma k|^\alpha \left[ 1 - i \beta \tan{\frac{\pi \alpha}{2}} sgn(k) \right]} & \alpha \neq 1 \\
        e^{-ik\delta}e^{-|\gamma k| \left[ 1 - i \frac{2}{\pi} \beta \log{|k|} sgn(k) \right]} & \alpha = 1
    \end{cases}  
\end{equation}
\end{widetext}

Where $\delta$ is a displacement parameter, $\gamma$ is a scale parameter (corresponding directly to a spatial width parameter for $\alpha=2$) with $\gamma \geq 0$, $\beta$ is a shape parameter analogous to skewness with $\beta \in [-1, 1]$, and $\alpha$ is a shape parameter called the characteristic exponent with $\alpha \in (0, 2]$. We note that special cases for Eqn.~\ref{eq:3.1} appear for $\alpha=1,~~\beta=0$, which is the Cauchy-Lorentz distribution; and the case $\alpha=2,~~\beta=0$, which is the Gaussian distribution.

Assuming $\hat{q}(k) = L\alpha S(k; \alpha, \gamma, \beta, \delta)$, and using Eqn.~\ref{eq:2.10}, we then find Eqn.~\ref{eq:3.2}. Going from Eqn.~\ref{eq:2.10} to Eqn.~\ref{eq:3.2} requires a Taylor series expansion of Eqn.~\ref{eq:3.1}, and then neglecting terms of order higher than $k^\alpha$. This is valid provided $|\hat{P}(k, t)| \ll \frac{1}{|\gamma k|^\alpha}$. As such, care should be taken for $\alpha \to 0$ and large $\gamma$.

\begin{equation}
\begin{split}
    \label{eq:3.2}
    \tau \frac{\partial \hat{P}}{\partial t} = & -\delta   i k \hat{P}(k,t) - \gamma^\alpha\hat{P}(k, t) |k|^\alpha \\ & + \beta \tan{\frac{\pi \alpha}{2}} \gamma^\alpha ik \hat{P}(k, t) |k|^{\alpha-1}
\end{split}
\end{equation}

To obtain an inverse transformation for this equation, especially given $\alpha \in \mathbb{R}^+$, we use the Reisz identity \cite{Balescu2007}:

\begin{equation}
    \label{eq:3.3}
    FT^{-1} \{ - |k|^\alpha\hat{f}(k) \} = D^a_{|x|} f(x)
\end{equation}

Where $D^\alpha_{|x|} f(x)$ is the (fractional) Reisz derivative, defined in Appendix~\ref{Appdx:A}. This allows us to write the fractional-advection, fractional-diffusion equation (or FADE): 

\begin{equation}
\begin{split}
    \label{eq:3.4}
    \tau \frac{\partial P}{\partial t} = & - \beta \tan{\frac{\pi \alpha}{2}} \gamma^\alpha \frac{\partial }{\partial x} \left( D^{\alpha - 1}_{|x|} P(x,t) \right) \\ & + \gamma^\alpha D^\alpha_{|x|} P(x,t) - \delta \frac{ \partial P(x,t)}{\partial x}
\end{split}
\end{equation}

Which is a non-local analogue of the advection-diffusion equation. The terms present on the RHS of Eqn.~\ref{eq:3.4} are, in order: the fractional advection term; the fractional diffusion term; and the standard advection term. In the limit $\alpha \to 2$, the non-locality vanishes and the equation becomes a standard advection-diffusion equation. 

The fractional advection term is so named by analogy to the advection term; it introduces anisotropy in transport, but unlike the advection term it is non-local in nature. As the FADE can describe both the local behaviour occurring for a Gaussian distribution and non-local behaviour arising from a L$\alpha$S distribution with $\alpha \neq 2$, then the FADE is a generalisation of the standard advection-diffusion relation which can be used to describe and quantify transport in general. 

\section{\label{sec:4} Jump Functions from STORM2D Simulations}
The STORM module of BOUT++ \cite{BOUTv4-3-2} solves 3D equations which are a drift-reduced, cold-ion, and electrostatic reduction of the Braginskii equations in the fluid limit. For this paper, we use STORM2D, which is a 2D reduced version of STORM and results in equations similar to those used by the ESEL code to model electrostatic interchange turbulence in the SOL \cite{Fundamenski2007}. Previous work with this package to simulate the SOL found that despite the simplifications made in the 2D model, there is overall good agreement with the 3D model in slab geometry in terms of the fluctuation statistics and radial density profiles \cite{Nicholas2021}.

The 2D equations result from grouping all terms representing transport parallel to the magnetic field into effective loss terms. This results in the following systems of transport equations for density, $n$, vorticity, $\Omega$, and temperature, $T$ \cite{Nicholas2021}:

\begin{equation}
    \label{eq:4.1}
    \frac{\partial n}{\partial t} = \frac{1}{B} \{ \phi, n \} + C(p) - n C(\phi) + D_n \nabla_{\perp}^2n -n_{loss} + S_{n0}
\end{equation}

\begin{equation}
    \label{eq:4.2}
    \frac{\partial \Omega}{\partial t} = \frac{1}{B} \{ \phi, \Omega \} + \frac{1}{n}C(p) + D_{\Omega}   \nabla_{\perp}^2 \Omega - \Omega_{loss} 
\end{equation}

\begin{equation}
\begin{split}
    \label{eq:4.3}
    \frac{\partial T}{\partial t} = & \frac{1}{B} \{ \phi, T \} - \frac{2}{3}TC(\phi) +  \frac{2}{3} \frac{T}{n} C(p) + \frac{5}{3}TC(T) \\ &  + \frac{2}{3} \frac{1}{n} \kappa_{\perp}\nabla_{\perp}^2T  + \frac{2}{3} \frac{1}{n} S_{E0} - \frac{1}{n} TS - T_{loss}
\end{split}
\end{equation}

and the vorticity related to the potential as: 

\begin{equation}
    \label{eq:4.4}
    \Omega = \nabla \cdot \left( \frac{\nabla_{\perp} \phi}{B^2} \right) 
\end{equation}

Where $C(g)$ denotes the curvature operator: 

\begin{equation}
    \label{eq:4.5}
    C(g) \equiv \nabla \times \frac{\textbf{b}}{B} \cdot \nabla g
\end{equation}

which may in simplified tokamak geometry at the midplane (which is indeed our case) be approximated as: 

\begin{equation}
    \label{eq:4.6}
    C(g) \approx - \frac{2}{R_0 B_0} \frac{\partial g}{\partial z}
\end{equation}

Where $\{a, b \}$ denotes the Poisson bracket in the x-z plane. There are two source terms: $S_n$ and $S_E$, corresponding to a density and energy source respectively. There are three sink/loss terms: $n_{loss}$, $\Omega_{loss}$, and $T_{loss}$, corresponding to density, vorticity and temperature loss terms respectively.

In this case, the geometry is assumed to be a tokamak flux tube with ($\textbf{B}$-) parallel, radial and binormal directions - in the simplified STORM2D equations here at the midplane, the parallel direction corresponds to the $y$-axis, the radial direction to the $x$-axis, and the binormal direction to the $z$-axis. The 2D equations are solved in the radial-binormal ($x-z$) direction, with the parallel terms approximated (hence the grouping of the parallel terms discussed earlier) as loss terms. 

For the set of simulations conducted, the parameters were selected as; $L_x=140~\rho_s$ with $1024$ cells, and $L_z=150~\rho_s$ with $256$ cells, and $L_y=L_\parallel=5500~\rho_s$. These particular settings were selected as they are similar to those used by Nicholas et al \cite{Nicholas2021}. The binormal boundary conditions are periodic, and the radial conditions are Dirichlet conditions. The three parallel loss terms $n_{loss}$, $\Omega_{loss}$ and $T_{loss}$, are defined in Appendix~\ref{Appdx:B}.

We use the Bohm normalisation as is usual. Spatial measures are normalised to the hybrid gyroradius $\rho_s = c_s/\Omega_i$, and measures of time to the ion gyrofrequency, $\Omega_i = eB_0/m_i$, where the normalising parameters are $m_i = 2~amu$, $B_0 = 0.25~T$, $n_0 = 0.8\times10^{19}~m^{-3}$, $T_0 = 40~eV$, $R_0=1.5~m$, $L_\parallel=5500~\rho_s$. The selected normalisations result in the velocity normalisation as $c_s = (T_0/m_i)^\frac{1}{2}$. Temperatures are normalised to some normalising temperature, $T_0$, densities normalised to $n_0$, transport coefficients normalised to the Bohm diffusion rate, $\rho_s^2 \Omega_i$, and the potential is normalised to $T_0/e$.

The transport parameters are calculated as in \cite{Fundamenski2007}, and the values are given for the reference case in table \ref{tab:4.1}. 

\begin{table}[]
    \centering
    \begin{tabular}{c|c|c|c|c}
    Param. & $D_n$ & $D_{\Omega}$ & $\kappa_{\perp}$ & $\kappa_{\parallel}$\\ 
    \hline
    Value  & $3.6 \times 10^{-3}$ & $7.1 \times 10^{-2}$ & $1.0 \times 10^{-2}$ & $1.1 \times 10^{5}$ \\
    \end{tabular}
    \caption{Transport parameters, reference case}
    \label{tab:4.1}
\end{table}

The simulations are set up with an initial density and energy source in order to imitate a density and energy source from the core plasma, which are localised to a narrow radial region. The region with the sources and radially within the sources represent a numerical buffer region. The region of the simulation radially outside the source regions is the part which we analyse, and which represents the SOL. The density and energy sources occur in the density and temperature transport equations (Eqns.~ \ref{eq:4.1} and \ref{eq:4.3}) as $S_n$ and $S_E$, and the sources have a form: 

\begin{equation}
    \label{eq:4.7}
    S_n = S_{n0} \frac{1}{\sqrt{2\pi w_{Sn}^2}} e^{-\frac{(x-x_{Sn})^2}{2w_{Sn}^2}}
\end{equation}
\begin{equation}
    \label{eq:4.8}
    S_E = S_{E0} \frac{1}{\sqrt{2\pi w_{SE}^2}} e^{-\frac{(x-x_{SE})^2}{2w_{SE}^2}}
\end{equation}

We have $x_{Sn}$ set such that the density source peaks $10\%$ into the domain, and $x_{SE}$ at $9\%$ of the domain. In both cases, the standard deviations $w$ are $0.25$, and for the reference case we set $S_{n0} = 0.03$ and $S_{E0} = 0.1$. Again, these are set similarly to Nicholas et al \cite{Nicholas2021}. We vary the magnitude of the density and energy sources around these parameter values such that we have a set of simulations with $S_{n0} \in [0.024, 0.042], S_{E0}=0.1$ and a set with $S_{E0} \in [0.08, 0.16], S_{n0}=0.03$. 

\begin{figure}
    \centering
    \includegraphics[width=8cm]{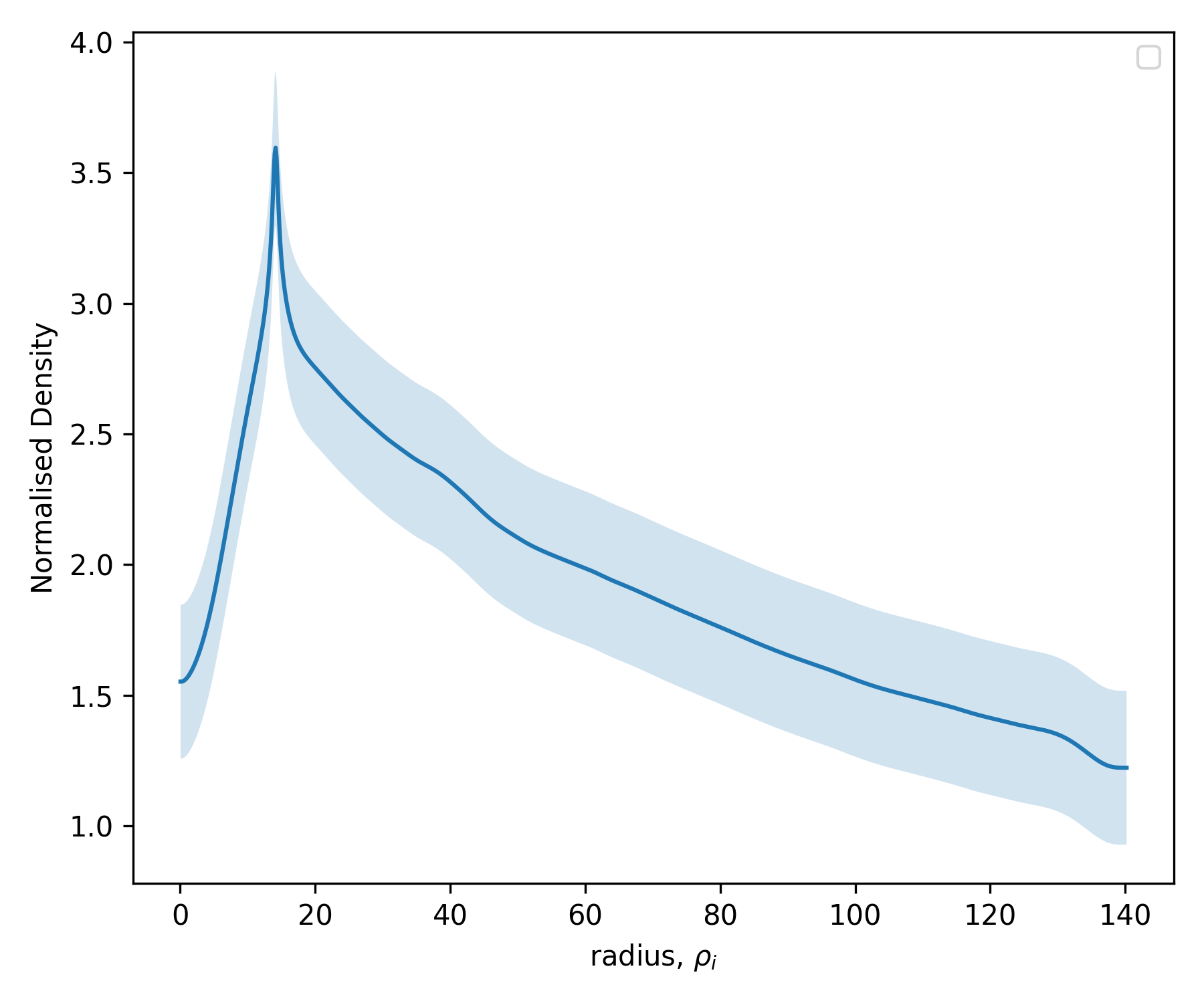}
    \caption{Normalised mean density generated by STORM for the reference case, with shaded region indicating the standard deviation}
    \label{fig:4.8b}
\end{figure}

\begin{figure*}
    \centering
    \includegraphics[width=16cm]{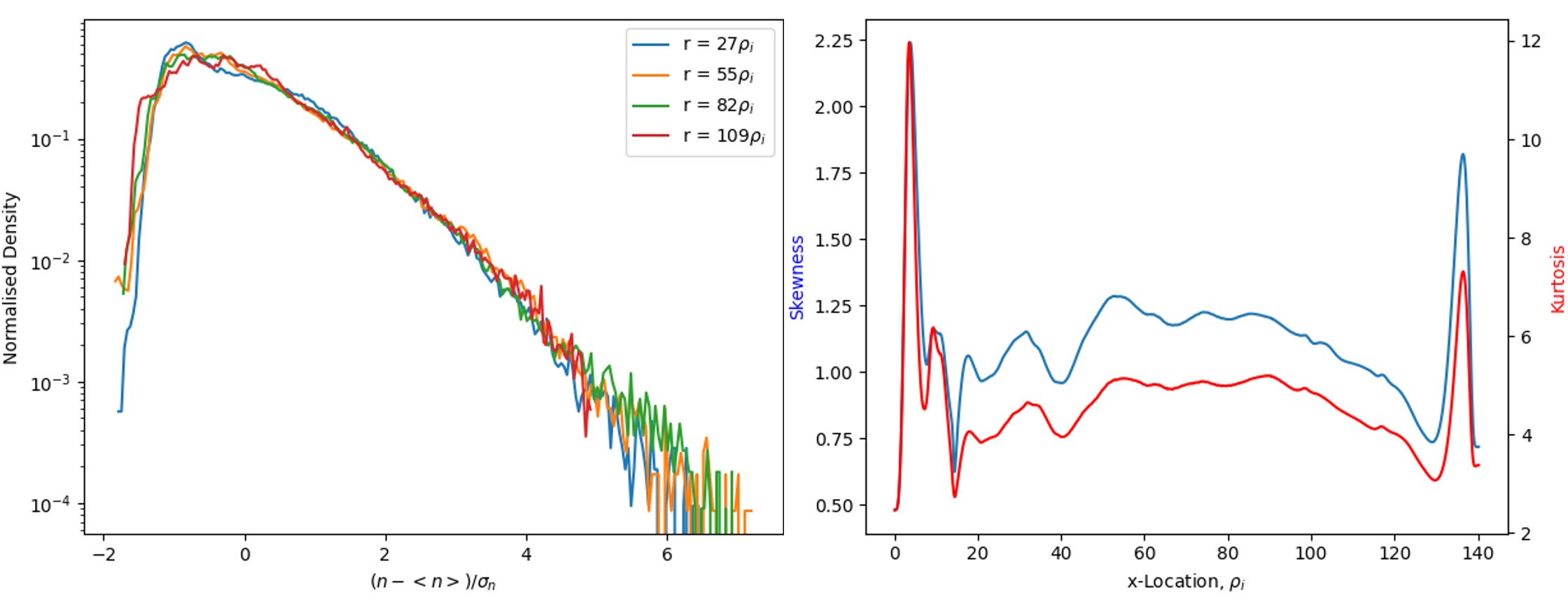}
    \caption{\textit{Left:} Density fluctuation PDF for a number of radii, basis case \textit{Right:} Skewness and Kurtosis of density fluctuations vs radius, basis case}
    \label{fig:4.8c}
\end{figure*}

We allow the simulation to run for a simulated time equivalent to several milliseconds to reach a steady-state turbulence: the statistical steady state acquired was used for analysis. Some properties of interest include the average profiles, as well as the fluctuation statistics. The average radial density profile for the base case are shown in Fig.~\ref{fig:4.8b}, and we show density fluctuation statistics in Fig.~\ref{fig:4.8c}. It should be noted that the simulations here reproduce the properties identified and discussed in Nicholas et al. \cite{Nicholas2021}, so a detailed discussion of these data is not provided. 

The stochastic flux equation and associated analysis can be used to show that a linear flux-gradient relation occurs if and only if we have a Gaussian jump function. In previous work \cite{Gheorghiu2024} we measured jump functions using Lagrangian tracers propagated by $\textbf{E} \times \textbf{B}$ drift, and demonstrated that in the case where zonal flows had formed in simulations of drift-wave turbulence using the Hasegawa-Wakatani equations, a non-Gaussian jump function was observed. In that case, tracers could be seeded uniformly across the domain. In the simulated SOL, due to the inhomogenous nature of the simulation and non-periodic radial boundaries the tracer seeding strategy is different, and consists of: defining a number of radial sectors exhaustively covering the domain in the radial direction in which tracers could be seeded; selecting an observation interval such that tracers in sectors next to the boundary sectors do not interfere with the radial boundaries, and specifying the widths of the radial sectors such that the observation interval is larger than the correlation time. On this basis we created 10 equally-sized radial sectors with an observation interval of $\tau = 5\times10^{-5}~s$ ($600$ periods of the ion gyrofrequency). To obtain good-quality jump functions with as minimal noise as possible, as many tracers must sample the space in as many configurations as possible. As such, the same simulations were restarted 10 times at different points in the steady-state turbulence regions, with jump functions gathered for the same radial positions. The datasets for the same radial sectors were then combined to provide a more representative picture. The data for each run then consisted of particle data from 10 different sectors, each with 10 restarts with $10^5$ particles per sector.

As we have just discussed, the identification of the jump function for density transport is dependent on the analysis of the motion of Lagrangian tracers which replicate the statistics of density transport. It is not immediately obvious how a particle tracking method could be used to identify a jump function for heat/energy transport. It seems likely that the FADE could be used to model heat/energy transport, but without the jump function, we do not have a way of determining the parameters from simulation - so we do not consider a FADE model of the transport of heat/energy in this paper. In Section~\ref{sec:X+1}, we discuss future work concerning this particular topic. 

\begin{figure*}
    \centering
    \includegraphics[width=16cm]{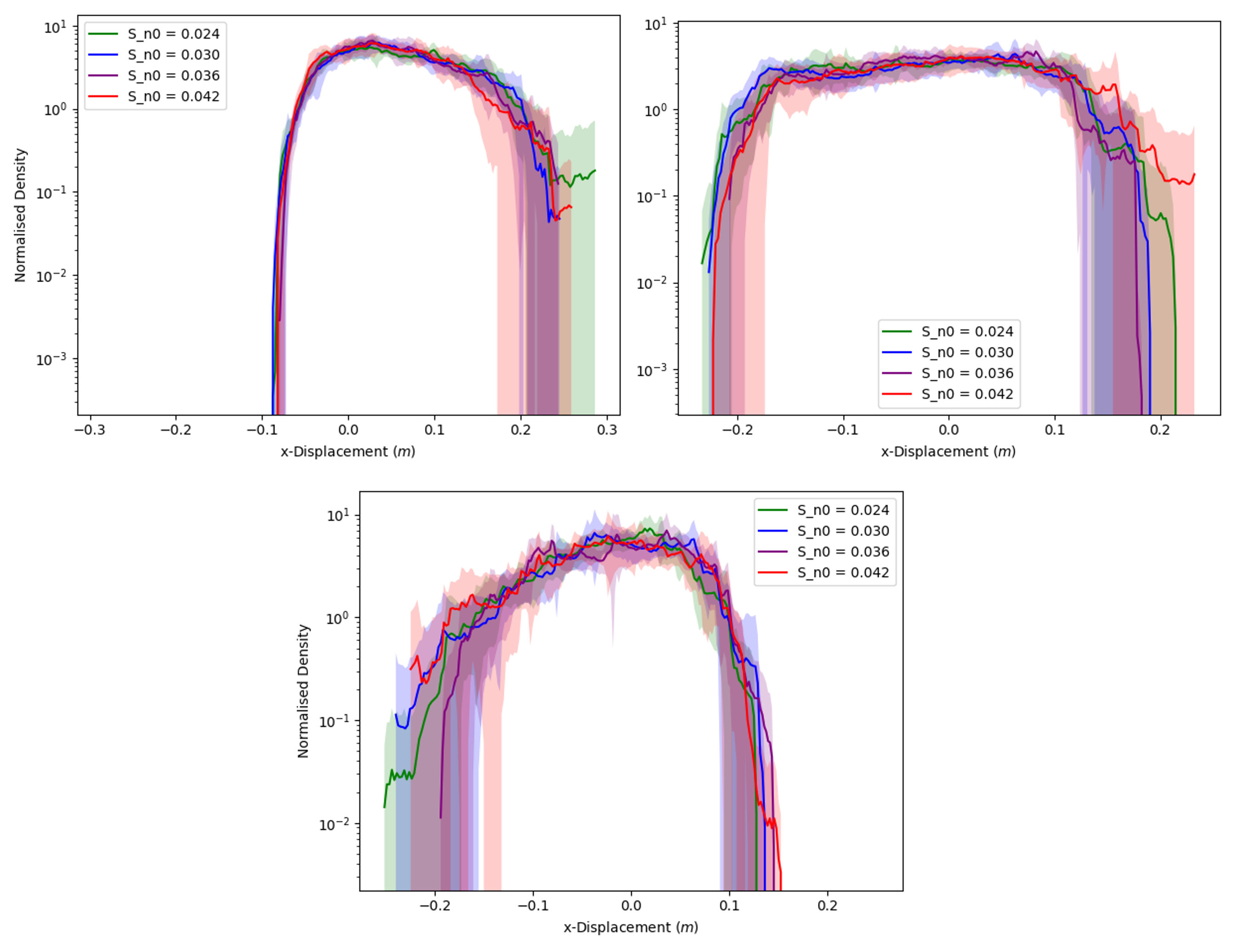}
    \caption{Ergodic jump functions for sectors $2$, $5$ and $7$ respectively (clockwise). The shaded areas denote the uncertainty in the data.}
    \label{fig:4.9}
\end{figure*}

\begin{figure*}
    \centering
    \includegraphics[width=16cm]{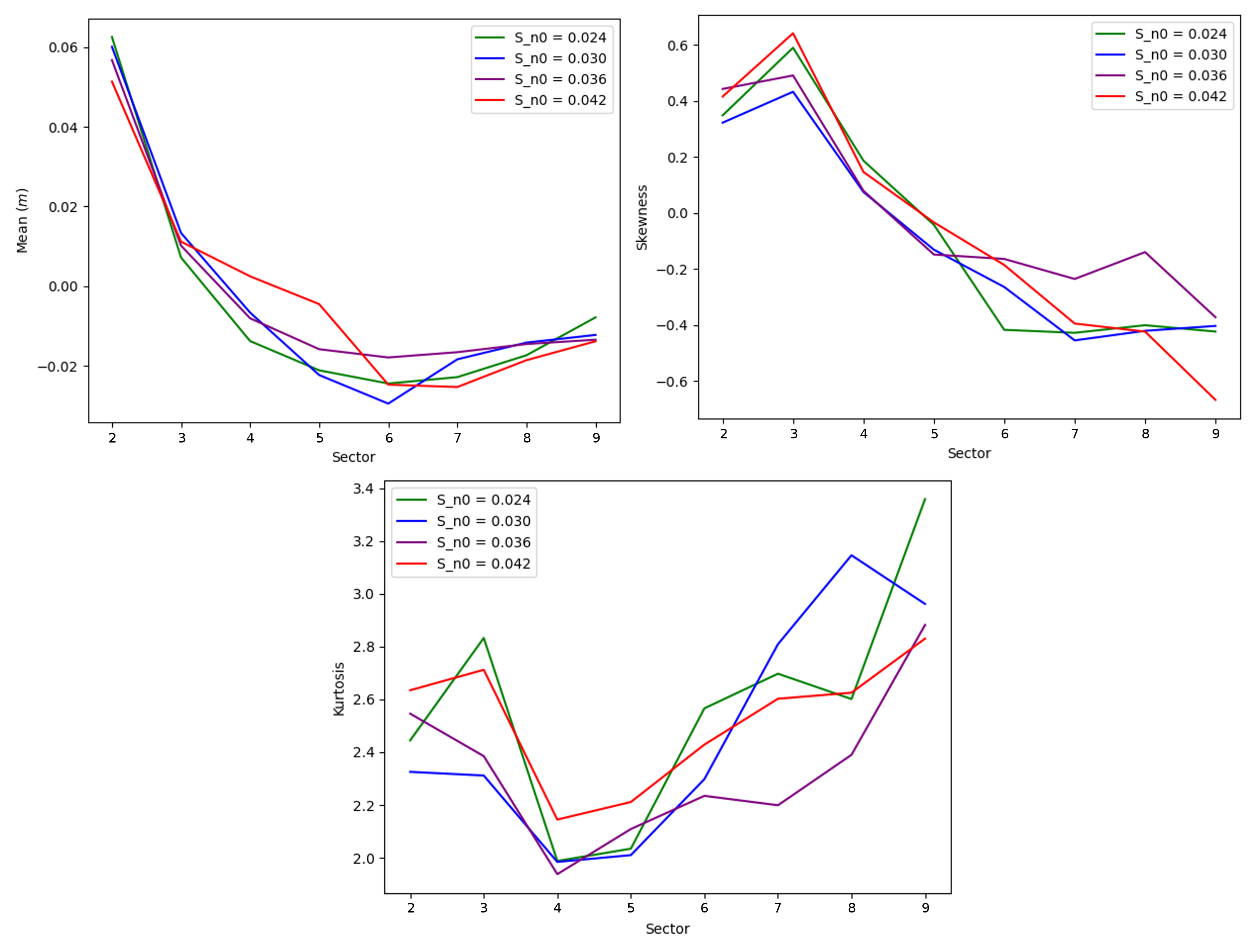}
    \caption{Variation in jump function statistics with sector for the varying density source runs - Mean, Skewness and Kurtosis respectively (clockwise). Note that sector refers to the splitting of the radial direction into 10 distinct regions}
    \label{fig:4.10}
\end{figure*}

Examples of the jump functions across the radius for the density source variation runs are shown in Fig.~\ref{fig:4.9}. The standard deviation is on the order of $20\%$ of the mean despite measures to improve the data quality. The jump functions do appear to be largely consistent across the change in the source magnitude. There appears to be a persistent change in the jump function properties across the radius, which strongly indicates that the jump function is radially inhomogeneous. The jump functions for the first and last sectors were discarded, as the data was influenced by the boundaries in an nonphysical manner: Additionally, the first sector was radially within the energy and density sources, and so not considered reflective of relevant physical conditions. We quantify this variation in terms of the mean, skewness and kurtosis of the jump functions with radius, which are shown in Fig.~\ref{fig:4.10}.

We notice an almost linear decrease in the skewness with radius from positive to negative values, and a reduction of the mean to negative values. Recall that the Gaussian distribution possesses a kurtosis equal to $3$, so the values identified suggest that the distributions are non-Gaussian. Very similar behaviour was observed for the scan in the energy source.

The Lagrangian tracer particle data suggests two things: First, the jump functions do not appear to be Gaussian, hence one cannot justify any kind of advection-diffusion equation as a reduced model; Second, the jump function appears to vary with radius, which may pose some difficulty as our analysis is based on a spatially uniform jump function. 

\begin{figure}
    \centering
    \includegraphics[width=8cm]{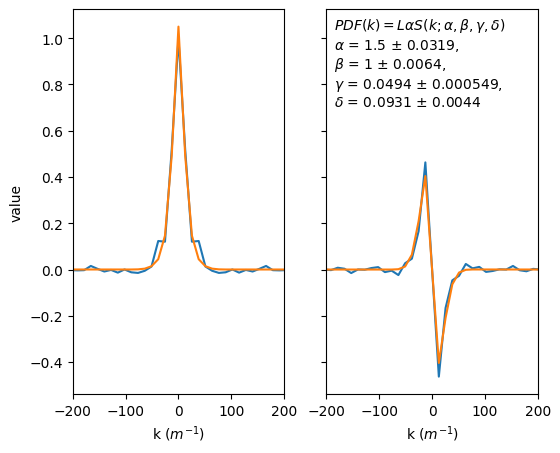}
    \caption{Fit in Fourier space of the L$\alpha$S distribution to the sector $2$ data for the reference case. Error in fit parameters are estimated from the covariance matrix of the fit. \textit{Left}: Real component of the fit. \textit{Right}: Imaginary component of the fit.}
    \label{fig:4.11}
\end{figure}

To mitigate this, we make some assumptions which leverage the physics of the system. The density is highest at the source region and decreases rapidly. When considering jump functions across the radius, the contribution of any particular jump function should be weighted by the local density: Consequently, we assume that the ensemble jump function would be primarily determined by the jump function(s) closer to the density peak - which in this case display a similar character. This difficulty does reveal a flaw with our Lagrangian tracers; ideally, there should be a way of weighting the particles or controlling their propagation that better reflects the actual system density - without requiring a full Particle-In-Cell simulation. 

As such, we consider the jump functions for sector $2$ and whether they are well-described by a L$\alpha$S distribution. We perform the fit in Fourier space as the L$\alpha$S is well-defined there. We give the example of the reference case to demonstrate the fitting, which are shown in Fig.~\ref{fig:4.11}: this shows that a L$\alpha$S distribution with $\alpha \approx1.5$ and $\beta\approx1$ is an appropriate fit. Note that, for example, sector $3$ demonstrates a higher skewness and so suggests a different fit. The jump functions from further out - e.g. sectors $7$ and $8$ - are very flat and do not correspond well to a stable distribution. However, overall non-integer $\alpha$ and non-zero $\beta$ indicate that a FADE may be an appropriate description of the dynamics in this case. 

We present the fits for the rest of the jump functions for the same sector across the $S_{n0}$ and $S_{e0}$ runs in table \ref{tab:4.2}. Overall trends include ubiquitously non-integer $\alpha$ in the range $[1.4, 1.7]$ and error of order $\approx 2\%$, a positive skewness with error of order $\approx 20\%$, a consistent $\gamma$ or `frequency width' with error of order $\approx 1\%$, and a non-zero displacement term $\delta$ with error of order $\approx 5\%$. The persistent non-zero $\delta$ suggests an advective contribution to the dynamics. 

\begin{table*}
    \centering
    \begin{tabular}{c||c|c|c|c}
    Run & $\alpha$ & $\beta$ & $\gamma$ & $\delta$\\ 
    \hline \hline
    $S_{n0}=0.03~S_{e0}=0.10~$ & $1.53 \pm 0.0335$ & $1.00 \pm 0.249$ & $0.0494 \pm 0.000542$ & $0.0891 \pm 0.00404$ \\
    $S_{n0}=0.024~S_{e0}=0.10$ & $1.60 \pm 0.0487$ & $1.00 \pm 0.388$ & $0.0546 \pm 0.000813$ & $0.0867 \pm 0.00499$ \\
    $S_{n0}=0.036~S_{e0}=0.10$ & $1.49 \pm 0.0301$ & $1.00 \pm 0.0219$ & $0.0473 \pm 0.000497$ & $0.0882 \pm 0.00402$ \\
    $S_{n0}=0.042~S_{e0}=0.10$ & $1.69 \pm 0.0402$ & $1.00 \pm 0.0585$ & $0.0481 \pm 0.000535$ & $0.0641 \pm 0.00273$ \\
    \hline \hline
    $S_{n0}=0.03~S_{e0}=0.10~$ & $1.53 \pm 0.0335$ & $1.00 \pm 0.249$ & $0.0494 \pm 0.000542$ & $0.0891 \pm 0.00404$ \\
    $S_{n0}=0.03~S_{e0}=0.08$ & $1.42 \pm 0.0335$ & $1.00 \pm 0.238$ & $0.0495 \pm 0.000638$ & $0.101 \pm 0.00626$ \\
    $S_{n0}=0.03~S_{e0}=0.12$ & $1.50 \pm 0.0335$ & $1.00 \pm 0.159$ & $0.0577 \pm 0.000903$ & $0.109 \pm 0.00728$ \\
    $S_{n0}=0.03~S_{e0}=0.14$ & $1.53 \pm 0.0511$ & $1.00 \pm 0.0351$ & $0.0588 \pm 0.000952$ & $0.108 \pm 0.00805$ \\
    $S_{n0}=0.03~S_{e0}=0.16$ & $1.54 \pm 0.0503$ & $1.00 \pm 0.0586$ & $0.0637 \pm 0.00103$ & $0.113 \pm 0.00739$ \\
    \end{tabular}
    \caption{Table of L$\alpha$S parameters for sector $2$ jump functions for various runs. The reference case is shown in the first row}
    \label{tab:4.2}
\end{table*}

\section{\label{sec:5} Comparison of mean profiles to solutions of FADEs}
The results of the preceding section indicate that we may attempt to model the density transport in the SOL using a FADE. The validity conditions specified for use of the FADE in section \ref{sec:3} were satisfied given the range of $\gamma$ and $\alpha$ and the constraints on the system size. Hence, we solve a 1D equation similar to Eqn.~\ref{eq:3.4}, of the form: 

\begin{equation}
\begin{split}
    \label{eq:5.1}
    \tau \frac{\partial n}{\partial t} = &  - \beta \tan{\frac{\pi \alpha}{2}} \gamma^\alpha \frac{\partial }{\partial x} \left( D^{\alpha - 1}_{|x|} n(x,t) \right) \\ & + \gamma^\alpha D^\alpha_{|x|} n(x,t)- \delta \frac{ \partial n(x,t)}{\partial x} - n_{loss} + S_n
\end{split}
\end{equation}

We solve this numerically by discretising in time using an explicit RK4 predictor-corrector scheme, and for the spatial discretisation we use the fractional Ortigueira operator \cite{Ortigueira2006} (see Appendix~\ref{Appdx:C}). We allow floating non-local boundary conditions in the ghost cells, which are updated based on the average of the nearest cells in the domain (See Appendix~\ref{Appdx:D}). This was done to ensure the closest replication of the STORM2D boundary conditions. We also add a density source at the same location as in the STORM2D simulations - this is located $10\%$ of the way into the domain, at approximately $14\rho_s$. The density and energy sources in the STORM2D simulations serve the role of a source from the `core' - so in the FADE simulations, we adjust the value of the density source to match the peak density in the SOL simulations. While it is feasible that the source region could have been modelled as some kind of inner Robin or nonlocal Robin boundary condition, the most straightforward approach is to consider the source as the `inner' boundary condition and match it in the FADE simulations. We apply the loss term only in the open field line region - that is in the SOL for radial values greater than the LCFS. Normalising the parameters is mostly straightforward using the convention used for STORM2D, if we recall that $\beta$ and $\alpha$ are already dimensionless. The observation interval is normalised as $\Omega_i\tau \to \tau$, the displacement parameter as $\delta/\rho_s \to \delta$, and the `frequency width' as $\gamma/\rho_s \to \gamma$. For the reference case then, we find that the normalised $\tau = 603$, the normalised $\gamma = 13.6$, and the normalised $\delta = 24.7$. As indicated in table \ref{tab:4.2}, we use $\alpha=1.53$ and $\beta=1$. 

\begin{figure}[h]
    \centering
    \includegraphics[width=8cm]{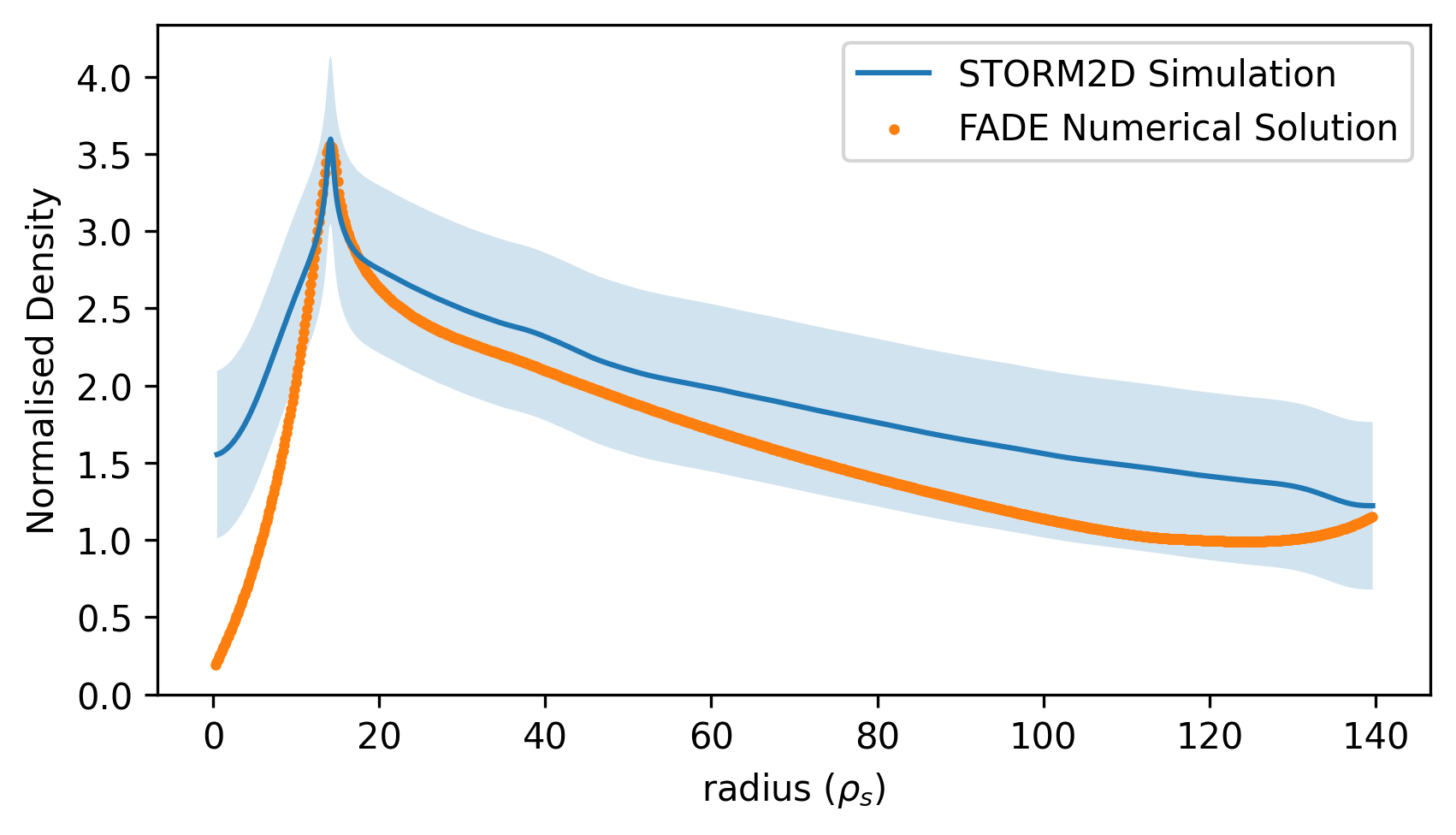}
    \caption{Comparison of mean density profile in reference case to converged FADE. The shaded blue region denotes the standard deviation in the density profile for the STORM simulations}
    \label{fig:5.1}
\end{figure}

The loss term is implemented in the solver as a proportional sink, with coefficients found using the estimated parameters. $L_{\parallel}$ is $5500$, and $V_{sh} \in [0.1, 1]$, so the magnitude of the loss term is estimated as being in the range $10^{-4} - 10^{-6}$. For these simulation, we use $10^{-5}$. For the particular parameters, Eqn. \ref{eq:5.1} converged well. The comparison of the converged density profile to the average density profile simulated by STORM2D is shown in Fig.~\ref{fig:5.1}. The converged FADE solution is a good match to the average profile from the STORM2D equations. Recall that the parameters used for the FADE are taken entirely from the sector $2$ jump function which we measured in the STORM2D simulation.

For the rest of the cases in table \ref{tab:4.2}, there was a similar match too. There is typically a good match between the FADE and the simulation density profiles, with the FADE typically lying in the uncertainty region for the STORM2D profile - i.e. within $\sim 20\%$ of the mean value. In Appendix~\ref{Appdx:E} we show a few scans over $\alpha$ and $\beta$ demonstrating that the FADE is highly sensitive to the equation parameters - as such, it seems the parameters identified in table \ref{tab:4.2} do reflect the system dynamics, and the good match is not merely a coincidence. Scans in the $\gamma$ parameter resulted in non-convergent FADEs. Note that the methods here can identify advective-diffusive, local, transport \textit{if present}: this would occur for the case $\alpha \to 2$. 

Given the match between the FADEs and the STORM2D simulations, this suggests that the approximations we made were reasonable, and it seems that we have evidence that supports the reduction of conventional transport equations to non-local FADEs in steady-state, and that a jump function can be defined and measured in order to identify said FADE parameters, at least in the regime studied here.

\section{\label{sec:X+1} Discussion and Conclusion} 
This study has demonstrated that density transport in the simulated SOL can be effectively described using fractional-advection, fractional-diffusion equations. By deriving the FADE parameters from the jump functions obtained through 2D simulations, we have established a correspondence between the FADE solutions and mean density profiles from the STORM2D simulations.

The next step is to develop the methods described here such that they may be applied to experimental data from the Tokamak SOL - this would permit experimental validation of the method, and thus permit characterisation of the SOL in terms of the FADE effective parameters. The main obstruction to applying the methods here to experimental data is the measurement of the jump function - it is not practical to track Lagrangian tracers in the SOL in the same manner as we have achieved here. However, it may be possible to infer jump functions indirectly using the flux-gradient relation - using Eqn.~\ref{eq:2.8}, the equation relating the profile to the flux, it is possible after some manipulation to directly relate the local gradient to the local flux - then, by assuming that the jump function belongs to the L$\alpha$S distribution, it should be possible to use a Bayesian method to infer the jump function parameters: This would therefore permit direct identification of FADE parameters without tracers - an approach similar to that in \cite{Pradalier2010}. 

Beyond this, there are numerous issues of interest, a few of which we detail here. While it is useful and interesting to measure jump functions, we should pay significant attention to the construction of jump functions from first principles. If we can do so based on our knowledge of the phenomena in the system of interest, this would allow us to create predictive non-local models of transport, which would be of significant use across the physical sciences. In the specific case of the SOL, we should direct our efforts to the construction of jump functions based on properties such as filament statistics or the statistics of turbulence. Another matter of interest, briefly discussed as limitations of the approach in section~\ref{sec:2}, is to investigate the use and application of spatially inhomogenous and time-varying jump functions in the stochastic flux formalism - this could potentially expand the range of application of the method significantly, and lead to other reduced transport equations. 

In this paper, we have considered primarily the application to density transport - however, the application to to other properties such as thermal energy/heat, which also show strong nonlocal behaviour \cite{Pradalier2010} would naturally also be of great interest. It is not immediately obvious how one could construct a `heat' jump function analogous to the density jump function, but this should be attempted nonetheless - regardless, it should still be possible to measure a jump function for other variables, using the flux-gradient approach proposed above. One final issue of interest is the possibility of correspondence between the reduced equations generated from the stochastic flux approach such as FADEs, and the full transport equations - if the equations are indeed equivalent in certain limits, it should be possible to rigorously demonstrate this; a general method to convert full transport equations to reduced non-local transport equations would be immensely useful. 

These brief considerations suggest that this is an area with a wide range of potential application: only by innovative and dedicated work in the realms of experiment and theory will the full scope become clear. 

\section*{Acknowledgements}
T.G. is grateful to John Omotani for many helpful discussions, as well as Stefan Mijin. This work was supported by the Engineering and Physical Sciences Research Council [EP/S022430/1]. This work has been carried out within the framework of the EUROfusion Consortium, funded by the European Union via the Euratom Research and Training Programme (Grant Agreement No 101052200 — EUROfusion) and from the EPSRC [grant number EP/W006839/1]. To obtain further information on the data and models underlying this paper please contact PublicationsManager@ukaea.uk. Views and opinions expressed are however those of the author(s) only and do not necessarily reflect those of the European Union or the European Commission. Neither the European Union nor the European Commission can be held responsible for them. This work was granted access to the HPC resources of EPCC via access to ARCHER2 by the Plasma HEC Consortium, EPSRC grant numbers EP/R029148/1 and EP/X035336/1.

\appendix
\section{\label{Appdx:A}}

The Reisz derivative is defined as follows \cite{Balescu2007}:

\begin{equation}
    \label{eq:A.1}
    D^\alpha_{|x|} f(x) = -\frac{1}{2 cos \frac{\pi \alpha}{2}} \left[ _{-\infty}D_x^\alpha + _x D_{\infty}^\alpha \right] f(x)
\end{equation}

And $_{-\infty}D_x^\alpha$, $_x D_{\infty}^\alpha$ denote the Left- and Right- fractional Riemann-Louiville Derivatives, where: 

\begin{equation}
    \label{eq:A.2}
     _{a}D_x^\alpha f(x) \equiv \frac{1}{\Gamma (m - \alpha)} \frac{d^m}{dx^m} \int^x_a \frac{f(\xi)}{(x - \xi)^{\alpha + 1 - m}} d \xi
\end{equation}

and:

\begin{equation}
    \label{eq:A.3}
     _x D_{b}^\alpha f(x) \equiv \frac{(-1)^m}{\Gamma (m - \alpha)} \frac{d^m}{dx^m} \int^b_x \frac{f(\xi)}{(\xi - x)^{\alpha + 1 - m}} d \xi
\end{equation}

Where in both cases $\alpha>0$ and $m-1\leq \alpha < m$, and $m \in \mathbf{N}$ - note that these conditions require $m$ to be the next integer larger than $\alpha$.

\section{\label{Appdx:B}}

The density and vorticity loss terms are determined by assuming a sheath dissipation closure \cite{Garcia2006b}, with: 

\begin{equation}
    \label{eq:B.1}
    n_{loss} = \frac{n}{L_{\parallel}} V_{sh}(\phi, T)
\end{equation}

\begin{equation}
    \label{eq:B.2}
    \Omega_{loss} = \frac{1}{L_{\parallel}} (V_{sh}(\phi, T) - \sqrt{T})
\end{equation}

Where $V_{sh}$ is the sheath velocity, given \cite{Nicholas2021} as: 

\begin{equation}
    \label{eq:B.3}
    V_{sh}(\phi, T) = V_{prefactor}\sqrt{T} e^{-\frac{\phi}{T}}
\end{equation}

Where $V_{prefactor}$ is the sheath prefactor, calculated as:

\begin{equation}
    \label{eq:B.4}
    V_{prefactor} \equiv \sqrt{\frac{\mu}{2\pi}\frac{\mu}{\mu + 1}}
\end{equation}

and $\mu \equiv \frac{m_i}{m_e}$. For the thermal loss term, we write: 

\begin{equation}
    \label{eq:B.5}
    T_{loss} = \frac{2}{3} \frac{1}{nL_{\parallel}} q_{\parallel}
\end{equation}

Where $q_{\parallel}$ is the harmonic average of the sheath-limited and conduction-limited parallel currents \cite{Myra2011}, given respectively as: 

\begin{equation}
    \label{eq:B.6}
    q_{\parallel} = (\gamma - \frac{3}{2})nTV_{sh}(\phi, T)
\end{equation}

\begin{equation}
    \label{eq:B.7}
    q_{\parallel} = \frac{2}{7}\frac{1}{L}\kappa_\parallel T^{\frac{7}{2}}
\end{equation}

Where $\gamma=5.5$.

\section{\label{Appdx:C}}

We would like to discretise the fractional derivative, such that we may write: 

\begin{equation}
    \label{eq:C.1}
    D^\alpha_{|x|} P(x,t) \simeq M^\alpha P_j
\end{equation}

Where $M^\alpha$ is a matrix of order $N \times N$ where $N$ is our number of grid-points, and $P_j$ is a column vector containing values of $P$ at each grid point $j \in N$. In this case, we take a finite difference approach. The discretisation of the fractional derivative can be found by using the Grünwald-Letnikov representation of the fractional derivative \cite{OldhamSpanier1974, Ortigueira2004}: 

\begin{equation}
    \label{eq:C.2}
    G^\alpha f(x) = \lim_{h \to 0^+} \frac{1}{h^\alpha} \sum^{\infty}_{k = 0}(-1)^k \binom{\alpha}{k} f(x - kh)
\end{equation}

Particularly useful is the theorem proved in references \cite{Ortigueira2004, Podlubny1998}, which establishes that the Grünwald-Letnikov derivative is equivalent to the generalised Cauchy formula for derivatives - this makes it clear that the Grünwald-Letnikov derivative may be related to the fractional (Reisz) derivative (Eqn. \ref{eq:3.3}): This property may be exploited to find a fractional (centred) difference discretisation of fractional operators \cite{Ortigueira2006}. This may be expressed in matrix form as follows:

\begin{equation}
    \label{eq:C.6}
    M^\alpha = \frac{1}{h^\alpha}
    \begin{bmatrix}
    \omega^\alpha_0 & \omega^\alpha_1 & \omega^\alpha_2 & \ldots & \omega^\alpha_{N-1} & \omega^\alpha_{N} \\
    \omega^\alpha_1 & \omega^\alpha_0 & \omega^\alpha_1 & \ldots & \omega^\alpha_{N-2} & \omega^\alpha_{N-1} \\
    \omega^\alpha_2 & \omega^\alpha_1 & \ddots & & & \vdots \\
    \vdots & \vdots &  & \ddots &  & \vdots \\
    \omega^\alpha_{N-1} & \omega^\alpha_{N-2} & \ldots  & \ldots & \omega^\alpha_0 & \omega^\alpha_1 \\
    \omega^\alpha_{N} & \omega^\alpha_{N-1} & \ldots & \ldots & \omega^\alpha_1 & \omega^\alpha_0 \\
    \end{bmatrix}
\end{equation}

Where $M^\alpha$ is symmetric, $M^\alpha = (M^\alpha)^T$, and $h$ denotes the spatial distance between grid-points. The values of $\omega^\alpha_k$ are given as: 

\begin{equation}
    \label{eq:C.7}
    \omega^\alpha_k = \frac{(-1)^k \Gamma(\alpha + 1) sgn(\cos{\frac{\alpha\pi}{2}})}{\Gamma(\frac{\alpha}{2} - k + 1)\Gamma(\frac{\alpha}{2} + k + 1)}
\end{equation}

This works for arbitrary $\alpha \in [0,2]$ (except 1) which is convenient, and one may be satisfied that this is the case by computing $M^\alpha$ for the second-order accurate central difference, which corresponds to $\alpha=2$, and observe that it returns the expected stencil. It was found in later work \cite{Celik2012} that this fractional centred difference approach is accurate to $O(h^2)$. 

Now, in our FADE we have a fractional `advection' term too, not simply the symmetric fractional diffusion as in Eqn. \ref{eq:C.1}:

\begin{equation}
    \label{eq:C.8}
    \frac{\partial }{\partial x} \left( D^{\alpha - 1}_{|x|} P(x,t) \right) \simeq A^{\alpha-1} P_j
\end{equation}

We may write the second-order accurate central difference approximation for the first-order derivative as: 

\begin{equation}
    \label{eq:C.9}
    f'_j(x) = \frac{1}{2h}
    \begin{bmatrix}
        \ldots & 0 & -1 & 0 & 1 & 0 & \ldots  
    \end{bmatrix}
    \begin{bmatrix}
        \vdots\\
        f_{j-2} \\
        f_{j-1}\\
        f_{j}\\
        f_{j+1}\\
        f_{j+2}\\
        \vdots \\
    \end{bmatrix}
\end{equation}

We may then represent the first-order difference as a tridiagonal $N\times N$ antisymmetric matrix, $F$, using Eqn. \ref{eq:C.9} (Except at the boundaries). We may then express $A^{\alpha-1}$ as: 

\begin{equation}
    \label{eq:C.10}
    A^{\alpha-1} = F M^{\alpha - 1}
\end{equation}

As $A^{\alpha-1}$ is the product of a symmetric and an antisymmetric matrix, it is therefore antisymmetric. $A^{\alpha-1}$ is the discretisation of the `fractional' advection component. 

We therefore have a discretisation for the fractional operators in our FADE, and so can solve it numerically.

\section{\label{Appdx:D}}

There is a question of appropriate boundary conditions in the case of non-local transport equations like FADEs, due to the non-locality intrinsic in the equations. Conventional differential operators, when discretized, result in finite contributions from a small number of nearby nodes. However, non-local operators such as those in the FADE have finite contributions from all nodes in the system. 
If we are simulating only part of a physical system, the boundary conditions we impose reflect the contributions and transport from and to the other parts of the system. With conventional transport equations, we can simply allow local, conventional boundary conditions - which then models local transport to and from the other parts of the system. With non-local transport equations, we then need to consider the possibility of non-local transport to and from other parts of the system. 
For example, if we wished to accurately simulate the evolution of a region around a perturbation in an infinite substance with a homogenous $L\alpha S$ jump-function with $\alpha \notin \mathbb{N}$ granting it non-local behaviour, we would have to account for non-local transport into the simulated area due to the non-local substance outside the domain, as well as non-local transport out of the domain. 

Consequently, for the purposes of this study we have devised a non-local Dirichlet boundary. This models the non-local flux into the domain, as if the substance beyond that boundary maintains a specified value with zero gradient, but is transported with the same jump function as the substance in the simulated area - essentially, a non-local infinite reservoir. 

In a 1D system, the computation of the fractional derivative at a grid point $j$ may be written as in Eqn.~\ref{eq:D.1}, where the system is bounded at the lower end by $p_-$ and at the upper end by $p_+$, and $\omega^\alpha$ are as in Eqn.~\ref{eq:C.6}.

\begin{widetext}
\begin{equation}
    \label{eq:D.1}
    D^\alpha_{|x|,j} f(x,t) =  \frac{1}{h^\alpha}  
    \begin{bmatrix}
       & \ldots & \omega^\alpha_{p_-+1} & |~\omega^\alpha_{p_-} & \ldots & \omega^\alpha_0 & \ldots & \omega^\alpha_{p_+}~| & \omega^\alpha_{p_++1}  & \ldots 
    \end{bmatrix}
    \begin{bmatrix}
        \vdots \\
        f_{j-p_--1} \\
        \rule{0.8cm}{0.5pt} \\
        f_{j-p_-} \\
        \vdots \\
        f_{j}\\
        \vdots \\
        f_{j+p_+} \\
        \rule{0.8cm}{0.5pt} \\
        f_{j+p_++1} \\
        \vdots \\
    \end{bmatrix}
\end{equation}
\end{widetext}

We may then separate the contributions according to their origin, in Eqn.~\ref{eq:D.2} - the first represents the contribution to $j$ from within the domain, and then the other terms represent the contribution to $j$ from outside the domain. 

\begin{equation}
    \label{eq:D.2}
    \begin{split}
    h^\alpha D^\alpha_{|x|,j} f(x,t) & =  \sum^{p_-,~p_+}_{k=0} \omega^\alpha_k f_{j \mp k} + \sum^N_{k=p_-+1} \omega^\alpha_k f_{j-k} \\ & + \sum^N_{k=p_++1} \omega^\alpha_k f_{j+k} 
    \end{split}
\end{equation}

The non-local Dirichlet boundary condition is then created by setting $f_{j-k}=\zeta_-$, for $k>p_-+1$, $f_{j+k} = \zeta_+$, for $k>p_++1$, where $\zeta_\mp$ are constants. This then permits:

\begin{equation}
    \label{eq:D.3}
    \begin{split}
    h^\alpha D^\alpha_{|x|,j} f(x,t) & =  \sum^{p_-,~p_+}_{k=0} \omega^\alpha_k f_{j \mp k} + \zeta_-\sum^N_{k=p_-+1} \omega^\alpha_k \\ & + \zeta_+\sum^N_{k=p_++1} \omega^\alpha_k
    \end{split}
\end{equation}

The summations in the last two terms are then truncated by selecting an $N$ such that $|\zeta_\mp  \omega^\alpha_n| \ll 1$ for $n > N$. 

\section{\label{Appdx:E}}

\begin{figure}
    \centering
    \includegraphics[width=8cm]{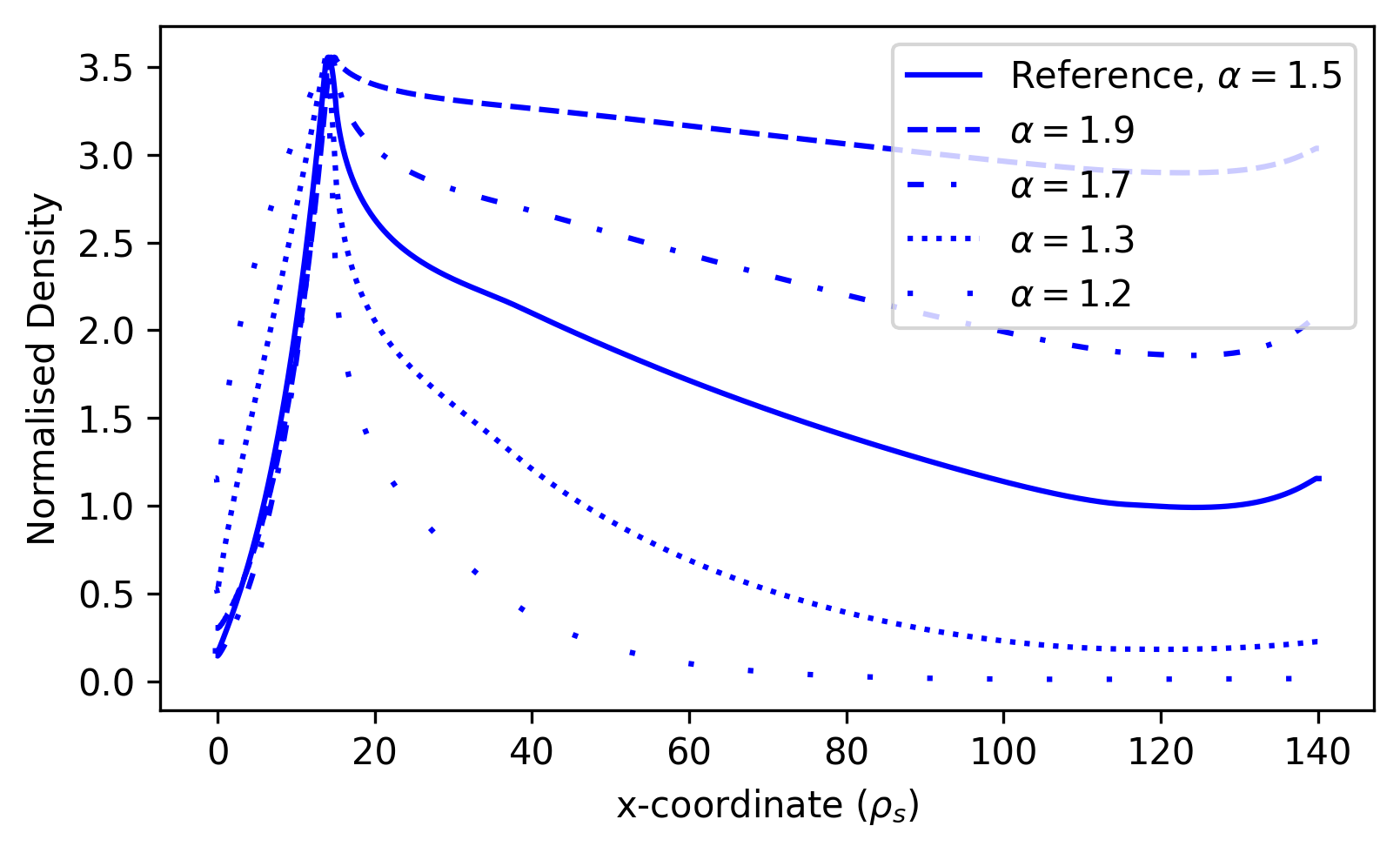}
    \caption{Scan of FADE in $\alpha$ around the parameters for the reference case ($\alpha = 1.5,~ \beta =1$)} 
    \label{fig:D.1}
\end{figure}

\begin{figure}
    \centering
    \includegraphics[width=8cm]{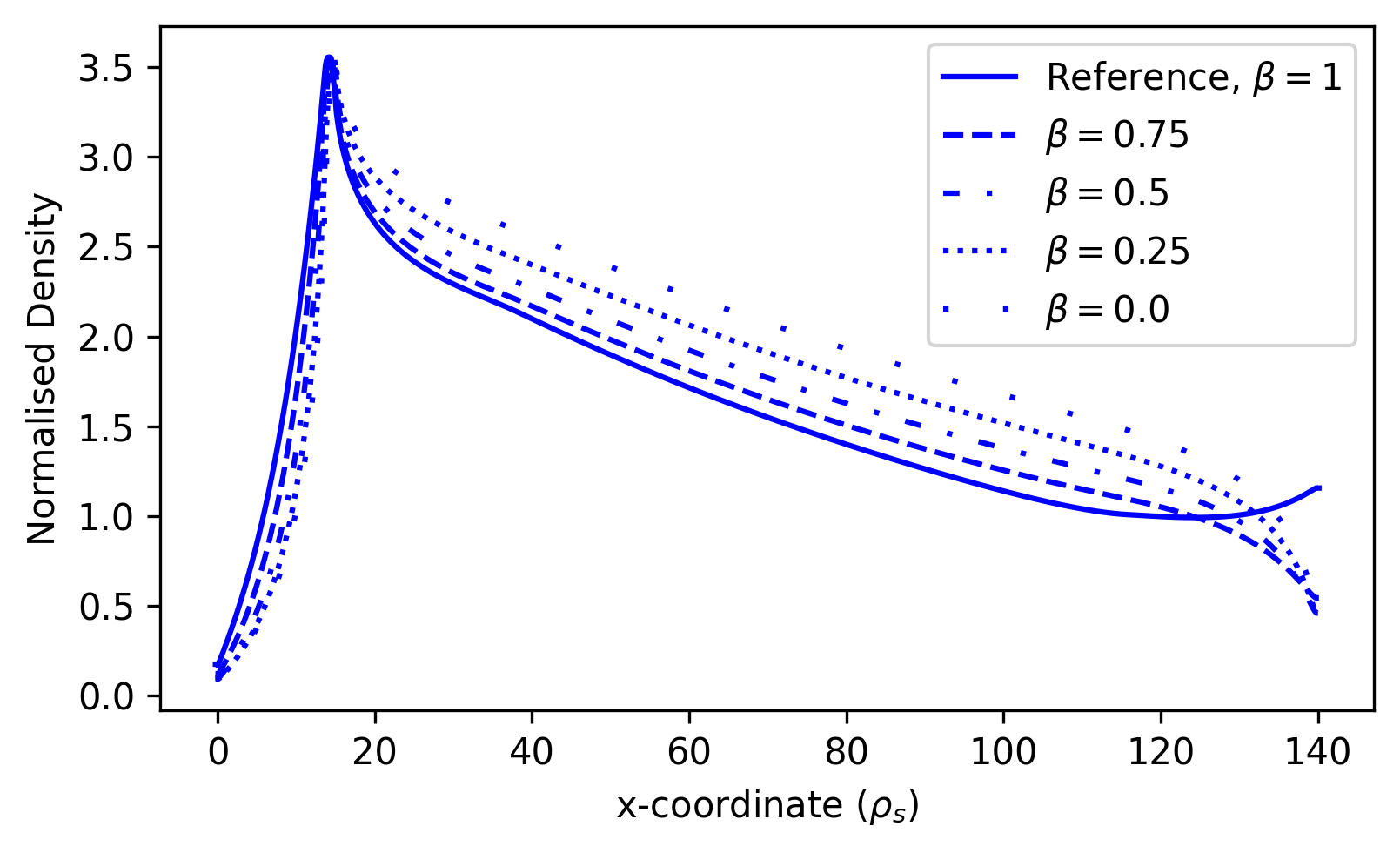}
    \caption{Scan of FADE in $\beta$ around the parameters for the reference case ($\alpha = 1.5,~ \beta =1$)}
    \label{fig:D.2}
\end{figure}

We show two scans - one in $\alpha$, one in $\beta$ - holding all other parameters the same as in the reference case, though altering the density source value to match the peak density. These may be seen in figures \ref{fig:D.1} and \ref{fig:D.2}. These scans demonstrate the significance of the characteristic exponent ($\alpha$) and the `skewness' ($\beta$) parameter in determining the properties of the solution. As can be seen the alteration of the characteristic exponent - the `nonlocal' parameter - has an extreme effect on the solution. Boundary conditions must be treated very carefully in nonlocal simulations - and this appears to be demonstrated in the $\beta$ scan for the right-hand boundary. Between $\beta=0.75$ and $\beta=1$, there is a strong change in the gradient of the profile after $\rho_s=120$. This is consistent with the observation that $\beta$ has the largest uncertainty in the fits. There is work to be done here ensuring the boundaries are appropriate, as this would eliminate one possible source of uncertainty. 

\newpage
\bibliography{apssamp}

\providecommand{\noopsort}[1]{}\providecommand{\singleletter}[1]{#1}%
\begin{thebibliography}{30}%
\makeatletter
\providecommand \@ifxundefined [1]{%
 \@ifx{#1\undefined}
}%
\providecommand \@ifnum [1]{%
 \ifnum #1\expandafter \@firstoftwo
 \else \expandafter \@secondoftwo
 \fi
}%
\providecommand \@ifx [1]{%
 \ifx #1\expandafter \@firstoftwo
 \else \expandafter \@secondoftwo
 \fi
}%
\providecommand \natexlab [1]{#1}%
\providecommand \enquote  [1]{``#1''}%
\providecommand \bibnamefont  [1]{#1}%
\providecommand \bibfnamefont [1]{#1}%
\providecommand \citenamefont [1]{#1}%
\providecommand \href@noop [0]{\@secondoftwo}%
\providecommand \href [0]{\begingroup \@sanitize@url \@href}%
\providecommand \@href[1]{\@@startlink{#1}\@@href}%
\providecommand \@@href[1]{\endgroup#1\@@endlink}%
\providecommand \@sanitize@url [0]{\catcode `\\12\catcode `\$12\catcode `\&12\catcode `\#12\catcode `\^12\catcode `\_12\catcode `\%12\relax}%
\providecommand \@@startlink[1]{}%
\providecommand \@@endlink[0]{}%
\providecommand \url  [0]{\begingroup\@sanitize@url \@url }%
\providecommand \@url [1]{\endgroup\@href {#1}{\urlprefix }}%
\providecommand \urlprefix  [0]{URL }%
\providecommand \Eprint [0]{\href }%
\providecommand \doibase [0]{https://doi.org/}%
\providecommand \selectlanguage [0]{\@gobble}%
\providecommand \bibinfo  [0]{\@secondoftwo}%
\providecommand \bibfield  [0]{\@secondoftwo}%
\providecommand \translation [1]{[#1]}%
\providecommand \BibitemOpen [0]{}%
\providecommand \bibitemStop [0]{}%
\providecommand \bibitemNoStop [0]{.\EOS\space}%
\providecommand \EOS [0]{\spacefactor3000\relax}%
\providecommand \BibitemShut  [1]{\csname bibitem#1\endcsname}%
\let\auto@bib@innerbib\@empty
\bibitem [{\citenamefont {Zweben}\ and\ \citenamefont {Gould}(1985)}]{Zweben1985}%
  \BibitemOpen
  \bibfield  {author} {\bibinfo {author} {\bibfnamefont {S.~J.}\ \bibnamefont {Zweben}}\ and\ \bibinfo {author} {\bibfnamefont {R.~W.}\ \bibnamefont {Gould}},\ }\bibfield  {title} {\bibinfo {title} {Structure of edge-plasma turbulence in the caltech tokamak structure of edge-plasma turbulence in the caltech tokamak},\ }\href@noop {} {\bibfield  {journal} {\bibinfo  {journal} {Nucl. Fusion}\ }\textbf {\bibinfo {volume} {25}},\ \bibinfo {pages} {171} (\bibinfo {year} {1985})}\BibitemShut {NoStop}%
\bibitem [{\citenamefont {Zweben}\ \emph {et~al.}(2004)\citenamefont {Zweben}, \citenamefont {Maqueda}, \citenamefont {Stotler}, \citenamefont {Keesee}, \citenamefont {Boedo}, \citenamefont {Bush}, \citenamefont {Kaye}, \citenamefont {LeBlanc}, \citenamefont {Lowrance}, \citenamefont {Mastrocola}, \citenamefont {Maingi}, \citenamefont {Nishino}, \citenamefont {Renda}, \citenamefont {Swain},\ and\ \citenamefont {Wilgen}}]{Zweben2004}%
  \BibitemOpen
  \bibfield  {author} {\bibinfo {author} {\bibfnamefont {S.~J.}\ \bibnamefont {Zweben}}, \bibinfo {author} {\bibfnamefont {R.~J.}\ \bibnamefont {Maqueda}}, \bibinfo {author} {\bibfnamefont {D.~P.}\ \bibnamefont {Stotler}}, \bibinfo {author} {\bibfnamefont {A.}~\bibnamefont {Keesee}}, \bibinfo {author} {\bibfnamefont {J.}~\bibnamefont {Boedo}}, \bibinfo {author} {\bibfnamefont {C.~E.}\ \bibnamefont {Bush}}, \bibinfo {author} {\bibfnamefont {S.~M.}\ \bibnamefont {Kaye}}, \bibinfo {author} {\bibfnamefont {B.}~\bibnamefont {LeBlanc}}, \bibinfo {author} {\bibfnamefont {J.~L.}\ \bibnamefont {Lowrance}}, \bibinfo {author} {\bibfnamefont {V.~J.}\ \bibnamefont {Mastrocola}}, \bibinfo {author} {\bibfnamefont {R.}~\bibnamefont {Maingi}}, \bibinfo {author} {\bibfnamefont {N.}~\bibnamefont {Nishino}}, \bibinfo {author} {\bibfnamefont {G.}~\bibnamefont {Renda}}, \bibinfo {author} {\bibfnamefont {D.~W.}\ \bibnamefont {Swain}},\ and\ \bibinfo {author} {\bibfnamefont {J.~B.}\ \bibnamefont {Wilgen}},\ }\bibfield  {title}
  {\bibinfo {title} {High-speed imaging of edge turbulence in nstx},\ }\href {https://doi.org/10.1088/0029-5515/44/1/016} {\bibfield  {journal} {\bibinfo  {journal} {Nuclear Fusion}\ }\textbf {\bibinfo {volume} {44}},\ \bibinfo {pages} {134} (\bibinfo {year} {2004})}\BibitemShut {NoStop}%
\bibitem [{\citenamefont {Garcia}\ \emph {et~al.}(2007{\natexlab{a}})\citenamefont {Garcia}, \citenamefont {Naulin}, \citenamefont {Nielsen},\ and\ \citenamefont {Rasmussen}}]{Garcia2007b}%
  \BibitemOpen
  \bibfield  {author} {\bibinfo {author} {\bibfnamefont {O.~E.}\ \bibnamefont {Garcia}}, \bibinfo {author} {\bibfnamefont {V.}~\bibnamefont {Naulin}}, \bibinfo {author} {\bibfnamefont {A.~H.}\ \bibnamefont {Nielsen}},\ and\ \bibinfo {author} {\bibfnamefont {J.~J.}\ \bibnamefont {Rasmussen}},\ }\bibfield  {title} {\bibinfo {title} {Turbulence and transport in the edge region of toroidally magnetized plasmas},\ }\href@noop {} {\bibfield  {journal} {\bibinfo  {journal} {Physics AUC}\ }\textbf {\bibinfo {volume} {17}},\ \bibinfo {pages} {263} (\bibinfo {year} {2007}{\natexlab{a}})}\BibitemShut {NoStop}%
\bibitem [{\citenamefont {Garcia}\ \emph {et~al.}(2007{\natexlab{b}})\citenamefont {Garcia}, \citenamefont {Pitts}, \citenamefont {Horacek}, \citenamefont {Nielsen}, \citenamefont {Fundamenski}, \citenamefont {Graves}, \citenamefont {Naulin},\ and\ \citenamefont {Rasmussen}}]{Garcia2007}%
  \BibitemOpen
  \bibfield  {author} {\bibinfo {author} {\bibfnamefont {O.~E.}\ \bibnamefont {Garcia}}, \bibinfo {author} {\bibfnamefont {R.~A.}\ \bibnamefont {Pitts}}, \bibinfo {author} {\bibfnamefont {J.}~\bibnamefont {Horacek}}, \bibinfo {author} {\bibfnamefont {A.~H.}\ \bibnamefont {Nielsen}}, \bibinfo {author} {\bibfnamefont {W.}~\bibnamefont {Fundamenski}}, \bibinfo {author} {\bibfnamefont {J.~P.}\ \bibnamefont {Graves}}, \bibinfo {author} {\bibfnamefont {V.}~\bibnamefont {Naulin}},\ and\ \bibinfo {author} {\bibfnamefont {J.~J.}\ \bibnamefont {Rasmussen}},\ }\bibfield  {title} {\bibinfo {title} {Turbulent transport in the tcv sol},\ }\href {https://doi.org/10.1016/j.jnucmat.2006.12.063} {\bibfield  {journal} {\bibinfo  {journal} {Journal of Nuclear Materials}\ }\textbf {\bibinfo {volume} {363-365}},\ \bibinfo {pages} {575} (\bibinfo {year} {2007}{\natexlab{b}})}\BibitemShut {NoStop}%
\bibitem [{\citenamefont {Naulin}(2007)}]{Naulin2007}%
  \BibitemOpen
  \bibfield  {author} {\bibinfo {author} {\bibfnamefont {V.}~\bibnamefont {Naulin}},\ }\bibfield  {title} {\bibinfo {title} {Turbulent transport and the plasma edge},\ }\href {https://doi.org/10.1016/j.jnucmat.2006.12.058} {\bibfield  {journal} {\bibinfo  {journal} {Journal of Nuclear Materials}\ }\textbf {\bibinfo {volume} {363-365}},\ \bibinfo {pages} {24} (\bibinfo {year} {2007})}\BibitemShut {NoStop}%
\bibitem [{\citenamefont {Yan}\ \emph {et~al.}(2021)\citenamefont {Yan}, \citenamefont {Chen}, \citenamefont {Xu}, \citenamefont {Wu}, \citenamefont {Liu}, \citenamefont {Wang}, \citenamefont {Meng}, \citenamefont {Hu}, \citenamefont {Zhao}, \citenamefont {Lan}, \citenamefont {Naulin}, \citenamefont {Nielsen}, \citenamefont {Rasmussen}, \citenamefont {Li}, \citenamefont {Wang}, \citenamefont {Yang}, \citenamefont {Li}, \citenamefont {Ye}, \citenamefont {Ding}, \citenamefont {Tao}, \citenamefont {Qian}, \citenamefont {Hou}, \citenamefont {Liu},\ and\ \citenamefont {Liu}}]{Yan2021}%
  \BibitemOpen
  \bibfield  {author} {\bibinfo {author} {\bibfnamefont {N.}~\bibnamefont {Yan}}, \bibinfo {author} {\bibfnamefont {L.}~\bibnamefont {Chen}}, \bibinfo {author} {\bibfnamefont {G.~S.}\ \bibnamefont {Xu}}, \bibinfo {author} {\bibfnamefont {X.~Q.}\ \bibnamefont {Wu}}, \bibinfo {author} {\bibfnamefont {S.~C.}\ \bibnamefont {Liu}}, \bibinfo {author} {\bibfnamefont {Y.~F.}\ \bibnamefont {Wang}}, \bibinfo {author} {\bibfnamefont {L.~Y.}\ \bibnamefont {Meng}}, \bibinfo {author} {\bibfnamefont {G.~H.}\ \bibnamefont {Hu}}, \bibinfo {author} {\bibfnamefont {N.}~\bibnamefont {Zhao}}, \bibinfo {author} {\bibfnamefont {H.}~\bibnamefont {Lan}}, \bibinfo {author} {\bibfnamefont {V.}~\bibnamefont {Naulin}}, \bibinfo {author} {\bibfnamefont {A.~H.}\ \bibnamefont {Nielsen}}, \bibinfo {author} {\bibfnamefont {J.~J.}\ \bibnamefont {Rasmussen}}, \bibinfo {author} {\bibfnamefont {K.~D.}\ \bibnamefont {Li}}, \bibinfo {author} {\bibfnamefont {L.}~\bibnamefont {Wang}}, \bibinfo {author} {\bibfnamefont {Q.~Q.}\ \bibnamefont {Yang}},
  \bibinfo {author} {\bibfnamefont {M.~H.}\ \bibnamefont {Li}}, \bibinfo {author} {\bibfnamefont {Y.}~\bibnamefont {Ye}}, \bibinfo {author} {\bibfnamefont {R.}~\bibnamefont {Ding}}, \bibinfo {author} {\bibfnamefont {Y.~Q.}\ \bibnamefont {Tao}}, \bibinfo {author} {\bibfnamefont {Y.~Z.}\ \bibnamefont {Qian}}, \bibinfo {author} {\bibfnamefont {J.~L.}\ \bibnamefont {Hou}}, \bibinfo {author} {\bibfnamefont {X.}~\bibnamefont {Liu}},\ and\ \bibinfo {author} {\bibfnamefont {J.~B.}\ \bibnamefont {Liu}},\ }\bibfield  {title} {\bibinfo {title} {Dependence of upstream sol density shoulder on divertor neutral pressure observed in l-mode and h-mode plasmas in the east superconducting tokamak},\ }\bibfield  {journal} {\bibinfo  {journal} {Nuclear Fusion}\ }\textbf {\bibinfo {volume} {61}},\ \href {https://doi.org/10.1088/1741-4326/abfe47} {10.1088/1741-4326/abfe47} (\bibinfo {year} {2021})\BibitemShut {NoStop}%
\bibitem [{\citenamefont {Long}\ \emph {et~al.}(2024)\citenamefont {Long}, \citenamefont {Diamond}, \citenamefont {Ke}, \citenamefont {Chen}, \citenamefont {Cao}, \citenamefont {Xu}, \citenamefont {Xu}, \citenamefont {Hong}, \citenamefont {Tian}, \citenamefont {Yuan}, \citenamefont {Liu}, \citenamefont {Yan}, \citenamefont {Yang}, \citenamefont {Shen}, \citenamefont {Guo}, \citenamefont {Wang}, \citenamefont {Nie}, \citenamefont {Wang}, \citenamefont {Hao}, \citenamefont {Wang}, \citenamefont {Chen}, \citenamefont {Pan}, \citenamefont {Li}, \citenamefont {Chen},\ and\ \citenamefont {Zhong}}]{Long2024}%
  \BibitemOpen
  \bibfield  {author} {\bibinfo {author} {\bibfnamefont {T.}~\bibnamefont {Long}}, \bibinfo {author} {\bibfnamefont {P.}~\bibnamefont {Diamond}}, \bibinfo {author} {\bibfnamefont {R.}~\bibnamefont {Ke}}, \bibinfo {author} {\bibfnamefont {Z.}~\bibnamefont {Chen}}, \bibinfo {author} {\bibfnamefont {M.}~\bibnamefont {Cao}}, \bibinfo {author} {\bibfnamefont {X.}~\bibnamefont {Xu}}, \bibinfo {author} {\bibfnamefont {M.}~\bibnamefont {Xu}}, \bibinfo {author} {\bibfnamefont {R.}~\bibnamefont {Hong}}, \bibinfo {author} {\bibfnamefont {W.}~\bibnamefont {Tian}}, \bibinfo {author} {\bibfnamefont {J.}~\bibnamefont {Yuan}}, \bibinfo {author} {\bibfnamefont {Y.}~\bibnamefont {Liu}}, \bibinfo {author} {\bibfnamefont {Q.}~\bibnamefont {Yan}}, \bibinfo {author} {\bibfnamefont {Q.}~\bibnamefont {Yang}}, \bibinfo {author} {\bibfnamefont {C.}~\bibnamefont {Shen}}, \bibinfo {author} {\bibfnamefont {W.}~\bibnamefont {Guo}}, \bibinfo {author} {\bibfnamefont {L.}~\bibnamefont {Wang}}, \bibinfo {author} {\bibfnamefont
  {L.}~\bibnamefont {Nie}}, \bibinfo {author} {\bibfnamefont {Z.}~\bibnamefont {Wang}}, \bibinfo {author} {\bibfnamefont {G.}~\bibnamefont {Hao}}, \bibinfo {author} {\bibfnamefont {N.}~\bibnamefont {Wang}}, \bibinfo {author} {\bibfnamefont {Z.}~\bibnamefont {Chen}}, \bibinfo {author} {\bibfnamefont {Y.}~\bibnamefont {Pan}}, \bibinfo {author} {\bibfnamefont {J.}~\bibnamefont {Li}}, \bibinfo {author} {\bibfnamefont {W.}~\bibnamefont {Chen}},\ and\ \bibinfo {author} {\bibfnamefont {W.}~\bibnamefont {Zhong}},\ }\bibfield  {title} {\bibinfo {title} {On how structures convey non-diffusive turbulence spreading},\ }\href {https://doi.org/10.1088/1741-4326/ad40c0} {\bibfield  {journal} {\bibinfo  {journal} {Nuclear Fusion}\ }\textbf {\bibinfo {volume} {64}},\ \bibinfo {pages} {064002} (\bibinfo {year} {2024})}\BibitemShut {NoStop}%
\bibitem [{\citenamefont {Garcia}(2012)}]{Garcia2012}%
  \BibitemOpen
  \bibfield  {author} {\bibinfo {author} {\bibfnamefont {O.~E.}\ \bibnamefont {Garcia}},\ }\bibfield  {title} {\bibinfo {title} {Stochastic modeling of intermittent scrape-off layer plasma fluctuations},\ }\bibfield  {journal} {\bibinfo  {journal} {Physical Review Letters}\ }\textbf {\bibinfo {volume} {108}},\ \href {https://doi.org/10.1103/PhysRevLett.108.265001} {10.1103/PhysRevLett.108.265001} (\bibinfo {year} {2012})\BibitemShut {NoStop}%
\bibitem [{\citenamefont {Losada}\ \emph {et~al.}(2023)\citenamefont {Losada}, \citenamefont {Theodorsen},\ and\ \citenamefont {Garcia}}]{Losada2022}%
  \BibitemOpen
  \bibfield  {author} {\bibinfo {author} {\bibfnamefont {J.~M.}\ \bibnamefont {Losada}}, \bibinfo {author} {\bibfnamefont {A.}~\bibnamefont {Theodorsen}},\ and\ \bibinfo {author} {\bibfnamefont {O.~E.}\ \bibnamefont {Garcia}},\ }\bibfield  {title} {\bibinfo {title} {Stochastic modeling of blob-like plasma filaments in the scrape-off layer: Theoretical foundation},\ }\href {https://doi.org/10.1063/5.0144885} {\bibfield  {journal} {\bibinfo  {journal} {Physics of Plasmas}\ }\textbf {\bibinfo {volume} {30}},\ \bibinfo {pages} {042518} (\bibinfo {year} {2023})}\BibitemShut {NoStop}%
\bibitem [{\citenamefont {Militello}\ and\ \citenamefont {Omotani}(2016)}]{Militello2016b}%
  \BibitemOpen
  \bibfield  {author} {\bibinfo {author} {\bibfnamefont {F.}~\bibnamefont {Militello}}\ and\ \bibinfo {author} {\bibfnamefont {J.~T.}\ \bibnamefont {Omotani}},\ }\bibfield  {title} {\bibinfo {title} {Scrape off layer profiles interpreted with filament dynamics},\ }\bibfield  {journal} {\bibinfo  {journal} {Nuclear Fusion}\ }\textbf {\bibinfo {volume} {56}},\ \href {https://doi.org/10.1088/0029-5515/56/10/104004} {10.1088/0029-5515/56/10/104004} (\bibinfo {year} {2016})\BibitemShut {NoStop}%
\bibitem [{\citenamefont {Gheorghiu}\ \emph {et~al.}(2024)\citenamefont {Gheorghiu}, \citenamefont {Militello},\ and\ \citenamefont {Rasmussen}}]{Gheorghiu2024}%
  \BibitemOpen
  \bibfield  {author} {\bibinfo {author} {\bibfnamefont {T.}~\bibnamefont {Gheorghiu}}, \bibinfo {author} {\bibfnamefont {F.}~\bibnamefont {Militello}},\ and\ \bibinfo {author} {\bibfnamefont {J.~J.}\ \bibnamefont {Rasmussen}},\ }\bibfield  {title} {\bibinfo {title} {On the transport of tracer particles in two-dimensional plasma edge turbulence},\ }\bibfield  {journal} {\bibinfo  {journal} {Physics of Plasmas}\ }\textbf {\bibinfo {volume} {31}},\ \href {https://doi.org/10.1063/5.0172484} {10.1063/5.0172484} (\bibinfo {year} {2024})\BibitemShut {NoStop}%
\bibitem [{\citenamefont {Gentle}\ \emph {et~al.}(1995)\citenamefont {Gentle}, \citenamefont {Bravenec}, \citenamefont {Cima}, \citenamefont {Gasquet}, \citenamefont {Hallock}, \citenamefont {Phillips}, \citenamefont {Ross}, \citenamefont {Rowan}, \citenamefont {Wootton}, \citenamefont {Crowley}, \citenamefont {Heard}, \citenamefont {Ouroua}, \citenamefont {Schoch},\ and\ \citenamefont {Watts}}]{Gentle1995}%
  \BibitemOpen
  \bibfield  {author} {\bibinfo {author} {\bibfnamefont {K.~W.}\ \bibnamefont {Gentle}}, \bibinfo {author} {\bibfnamefont {R.~V.}\ \bibnamefont {Bravenec}}, \bibinfo {author} {\bibfnamefont {G.}~\bibnamefont {Cima}}, \bibinfo {author} {\bibfnamefont {H.}~\bibnamefont {Gasquet}}, \bibinfo {author} {\bibfnamefont {G.~A.}\ \bibnamefont {Hallock}}, \bibinfo {author} {\bibfnamefont {P.~E.}\ \bibnamefont {Phillips}}, \bibinfo {author} {\bibfnamefont {D.~W.}\ \bibnamefont {Ross}}, \bibinfo {author} {\bibfnamefont {W.~L.}\ \bibnamefont {Rowan}}, \bibinfo {author} {\bibfnamefont {A.~J.}\ \bibnamefont {Wootton}}, \bibinfo {author} {\bibfnamefont {T.~P.}\ \bibnamefont {Crowley}}, \bibinfo {author} {\bibfnamefont {J.}~\bibnamefont {Heard}}, \bibinfo {author} {\bibfnamefont {A.}~\bibnamefont {Ouroua}}, \bibinfo {author} {\bibfnamefont {P.~M.}\ \bibnamefont {Schoch}},\ and\ \bibinfo {author} {\bibfnamefont {C.}~\bibnamefont {Watts}},\ }\bibfield  {title} {\bibinfo {title} {An experimental counter-example to the local
  transport paradigm},\ }\href {https://doi.org/10.1063/1.871252} {\bibfield  {journal} {\bibinfo  {journal} {Physics of Plasmas}\ }\textbf {\bibinfo {volume} {2}},\ \bibinfo {pages} {2292} (\bibinfo {year} {1995})}\BibitemShut {NoStop}%
\bibitem [{\citenamefont {Dif-Pradalier}\ \emph {et~al.}(2010)\citenamefont {Dif-Pradalier}, \citenamefont {Diamond}, \citenamefont {Grandgirard}, \citenamefont {Sarazin}, \citenamefont {Abiteboul}, \citenamefont {Garbet}, \citenamefont {Ghendrih}, \citenamefont {Strugarek}, \citenamefont {Ku},\ and\ \citenamefont {Chang}}]{Pradalier2010}%
  \BibitemOpen
  \bibfield  {author} {\bibinfo {author} {\bibfnamefont {G.}~\bibnamefont {Dif-Pradalier}}, \bibinfo {author} {\bibfnamefont {P.~H.}\ \bibnamefont {Diamond}}, \bibinfo {author} {\bibfnamefont {V.}~\bibnamefont {Grandgirard}}, \bibinfo {author} {\bibfnamefont {Y.}~\bibnamefont {Sarazin}}, \bibinfo {author} {\bibfnamefont {J.}~\bibnamefont {Abiteboul}}, \bibinfo {author} {\bibfnamefont {X.}~\bibnamefont {Garbet}}, \bibinfo {author} {\bibfnamefont {P.}~\bibnamefont {Ghendrih}}, \bibinfo {author} {\bibfnamefont {A.}~\bibnamefont {Strugarek}}, \bibinfo {author} {\bibfnamefont {S.}~\bibnamefont {Ku}},\ and\ \bibinfo {author} {\bibfnamefont {C.~S.}\ \bibnamefont {Chang}},\ }\bibfield  {title} {\bibinfo {title} {On the validity of the local diffusive paradigm in turbulent plasma transport},\ }\bibfield  {journal} {\bibinfo  {journal} {Physical Review E - Statistical, Nonlinear, and Soft Matter Physics}\ }\textbf {\bibinfo {volume} {82}},\ \href {https://doi.org/10.1103/PhysRevE.82.025401} {10.1103/PhysRevE.82.025401}
  (\bibinfo {year} {2010})\BibitemShut {NoStop}%
\bibitem [{\citenamefont {Ida}(2022)}]{Ida2022}%
  \BibitemOpen
  \bibfield  {author} {\bibinfo {author} {\bibfnamefont {K.}~\bibnamefont {Ida}},\ }\bibfield  {title} {\bibinfo {title} {Non-local transport nature revealed by the research in transient phenomena of toroidal plasma},\ }\bibfield  {journal} {\bibinfo  {journal} {Reviews of Modern Plasma Physics}\ }\textbf {\bibinfo {volume} {6}},\ \href {https://doi.org/10.1007/s41614-022-00064-6} {10.1007/s41614-022-00064-6} (\bibinfo {year} {2022})\BibitemShut {NoStop}%
\bibitem [{\citenamefont {Del-Castillo-Negrete}(2006)}]{CastilloNegrete2006}%
  \BibitemOpen
  \bibfield  {author} {\bibinfo {author} {\bibfnamefont {D.}~\bibnamefont {Del-Castillo-Negrete}},\ }\bibfield  {title} {\bibinfo {title} {Fractional diffusion models of nonlocal transport},\ }\bibfield  {journal} {\bibinfo  {journal} {Physics of Plasmas}\ }\textbf {\bibinfo {volume} {13}},\ \href {https://doi.org/10.1063/1.2336114} {10.1063/1.2336114} (\bibinfo {year} {2006})\BibitemShut {NoStop}%
\bibitem [{\citenamefont {Del-Castillo-Negrete}\ \emph {et~al.}(2008)\citenamefont {Del-Castillo-Negrete}, \citenamefont {Mantica}, \citenamefont {Naulin},\ and\ \citenamefont {Rasmussen}}]{CastilloNegrete2008}%
  \BibitemOpen
  \bibfield  {author} {\bibinfo {author} {\bibfnamefont {D.}~\bibnamefont {Del-Castillo-Negrete}}, \bibinfo {author} {\bibfnamefont {P.}~\bibnamefont {Mantica}}, \bibinfo {author} {\bibfnamefont {V.}~\bibnamefont {Naulin}},\ and\ \bibinfo {author} {\bibfnamefont {J.~J.}\ \bibnamefont {Rasmussen}},\ }\bibfield  {title} {\bibinfo {title} {Fractional diffusion models of non-local perturbative transport: Numerical results and application to jet experiments},\ }\bibfield  {journal} {\bibinfo  {journal} {Nuclear Fusion}\ }\textbf {\bibinfo {volume} {48}},\ \href {https://doi.org/10.1088/0029-5515/48/7/075009} {10.1088/0029-5515/48/7/075009} (\bibinfo {year} {2008})\BibitemShut {NoStop}%
\bibitem [{\citenamefont {Dudson}\ \emph {et~al.}(2020)\citenamefont {Dudson}, \citenamefont {Hill}, \citenamefont {Dickinson}, \citenamefont {Parker}, \citenamefont {Dempsey},\ and\ \citenamefont {et~al.}}]{BOUTv4-3-2}%
  \BibitemOpen
  \bibfield  {author} {\bibinfo {author} {\bibfnamefont {B.~D.}\ \bibnamefont {Dudson}}, \bibinfo {author} {\bibfnamefont {P.~A.}\ \bibnamefont {Hill}}, \bibinfo {author} {\bibfnamefont {D.}~\bibnamefont {Dickinson}}, \bibinfo {author} {\bibfnamefont {J.}~\bibnamefont {Parker}}, \bibinfo {author} {\bibfnamefont {A.}~\bibnamefont {Dempsey}},\ and\ \bibinfo {author} {\bibnamefont {et~al.}},\ }\href {https://doi.org/10.5281/zenodo.4046792} {\bibinfo {title} {{BOUT++}}},\ \bibinfo {howpublished} {\url{https://github.com/boutproject/BOUT-dev}} (\bibinfo {year} {2020})\BibitemShut {NoStop}%
\bibitem [{\citenamefont {Nicholas}\ \emph {et~al.}(2022)\citenamefont {Nicholas}, \citenamefont {Omotani}, \citenamefont {Riva}, \citenamefont {Militello},\ and\ \citenamefont {Dudson}}]{Nicholas2021}%
  \BibitemOpen
  \bibfield  {author} {\bibinfo {author} {\bibfnamefont {T.~E.~G.}\ \bibnamefont {Nicholas}}, \bibinfo {author} {\bibfnamefont {J.}~\bibnamefont {Omotani}}, \bibinfo {author} {\bibfnamefont {F.}~\bibnamefont {Riva}}, \bibinfo {author} {\bibfnamefont {F.}~\bibnamefont {Militello}},\ and\ \bibinfo {author} {\bibfnamefont {B.}~\bibnamefont {Dudson}},\ }\bibfield  {title} {\bibinfo {title} {Comparing two- and three-dimensional models of scrape-off-layer turbulent transport},\ }\href {http://arxiv.org/abs/2103.09727} {\bibfield  {journal} {\bibinfo  {journal} {Plasma Physics and Controlled Fusion}\ }\textbf {\bibinfo {volume} {64}} (\bibinfo {year} {2022})}\BibitemShut {NoStop}%
\bibitem [{\citenamefont {Gheorghiu}(2025)}]{Gheorghiu2024b}%
  \BibitemOpen
  \bibfield  {author} {\bibinfo {author} {\bibfnamefont {T.~E.}\ \bibnamefont {Gheorghiu}},\ }\emph {\bibinfo {title} {Non-local Modelling of Radial Transport in the Tokamak Edge and SOL}},\ \href@noop {} {Ph.D. thesis},\ \bibinfo  {school} {University of York} (\bibinfo {year} {2025})\BibitemShut {NoStop}%
\bibitem [{\citenamefont {Nolan}(2020)}]{Nolan2020}%
  \BibitemOpen
  \bibfield  {author} {\bibinfo {author} {\bibfnamefont {J.~P.}\ \bibnamefont {Nolan}},\ }\href {https://doi.org/https://doi.org/10.1007/978-3-030-52915-4} {\emph {\bibinfo {title} {Univariate Stable Distributions}}},\ \bibinfo {edition} {first edition}\ ed.\ (\bibinfo  {publisher} {Springer},\ \bibinfo {year} {2020})\BibitemShut {NoStop}%
\bibitem [{\citenamefont {Gnedenko}\ and\ \citenamefont {Kolmogorov}(1952)}]{Gnedenko1952}%
  \BibitemOpen
  \bibfield  {author} {\bibinfo {author} {\bibfnamefont {B.~V.}\ \bibnamefont {Gnedenko}}\ and\ \bibinfo {author} {\bibfnamefont {A.~N.}\ \bibnamefont {Kolmogorov}},\ }\href@noop {} {\emph {\bibinfo {title} {Limit Distributions For Sums of Independent Random Variables}}},\ \bibinfo {edition} {1st}\ ed.,\ edited by\ \bibinfo {editor} {\bibfnamefont {E.}~\bibnamefont {Reissner}}\ (\bibinfo  {publisher} {Addison-Wesley Publishing},\ \bibinfo {year} {1952})\BibitemShut {NoStop}%
\bibitem [{\citenamefont {Balescu}(2007)}]{Balescu2007}%
  \BibitemOpen
  \bibfield  {author} {\bibinfo {author} {\bibfnamefont {R.}~\bibnamefont {Balescu}},\ }\bibfield  {title} {\bibinfo {title} {V-langevin equations, continuous time random walks and fractional diffusion},\ }\href {https://doi.org/10.1016/j.chaos.2007.01.050} {\bibfield  {journal} {\bibinfo  {journal} {Chaos, Solitons and Fractals}\ }\textbf {\bibinfo {volume} {34}},\ \bibinfo {pages} {62} (\bibinfo {year} {2007})}\BibitemShut {NoStop}%
\bibitem [{\citenamefont {Fundamenski}\ \emph {et~al.}(2007)\citenamefont {Fundamenski}, \citenamefont {Garcia}, \citenamefont {Naulin}, \citenamefont {Pitts}, \citenamefont {Nielsen}, \citenamefont {Rasmussen}, \citenamefont {Horacek},\ and\ \citenamefont {Graves}}]{Fundamenski2007}%
  \BibitemOpen
  \bibfield  {author} {\bibinfo {author} {\bibfnamefont {W.}~\bibnamefont {Fundamenski}}, \bibinfo {author} {\bibfnamefont {O.~E.}\ \bibnamefont {Garcia}}, \bibinfo {author} {\bibfnamefont {V.}~\bibnamefont {Naulin}}, \bibinfo {author} {\bibfnamefont {R.~A.}\ \bibnamefont {Pitts}}, \bibinfo {author} {\bibfnamefont {A.~H.}\ \bibnamefont {Nielsen}}, \bibinfo {author} {\bibfnamefont {J.~J.}\ \bibnamefont {Rasmussen}}, \bibinfo {author} {\bibfnamefont {J.}~\bibnamefont {Horacek}},\ and\ \bibinfo {author} {\bibfnamefont {J.~P.}\ \bibnamefont {Graves}},\ }\bibfield  {title} {\bibinfo {title} {Dissipative processes in interchange driven scrape-off layer turbulence},\ }\href {https://doi.org/10.1088/0029-5515/47/5/006} {\bibfield  {journal} {\bibinfo  {journal} {Nuclear Fusion}\ }\textbf {\bibinfo {volume} {47}},\ \bibinfo {pages} {417} (\bibinfo {year} {2007})}\BibitemShut {NoStop}%
\bibitem [{\citenamefont {Ortigueira}(2006)}]{Ortigueira2006}%
  \BibitemOpen
  \bibfield  {author} {\bibinfo {author} {\bibfnamefont {M.~D.}\ \bibnamefont {Ortigueira}},\ }\bibfield  {title} {\bibinfo {title} {Riesz potential operators and inverses via fractional centred derivatives},\ }\bibfield  {journal} {\bibinfo  {journal} {International Journal of Mathematics and Mathematical Sciences}\ }\textbf {\bibinfo {volume} {2006}},\ \href {https://doi.org/10.1155/IJMMS/2006/48391} {10.1155/IJMMS/2006/48391} (\bibinfo {year} {2006})\BibitemShut {NoStop}%
\bibitem [{\citenamefont {Garcia}\ \emph {et~al.}(2006)\citenamefont {Garcia}, \citenamefont {Bian},\ and\ \citenamefont {Fundamenski}}]{Garcia2006b}%
  \BibitemOpen
  \bibfield  {author} {\bibinfo {author} {\bibfnamefont {O.~E.}\ \bibnamefont {Garcia}}, \bibinfo {author} {\bibfnamefont {N.~H.}\ \bibnamefont {Bian}},\ and\ \bibinfo {author} {\bibfnamefont {W.}~\bibnamefont {Fundamenski}},\ }\bibfield  {title} {\bibinfo {title} {Radial interchange motions of plasma filaments},\ }\bibfield  {journal} {\bibinfo  {journal} {Physics of Plasmas}\ }\textbf {\bibinfo {volume} {13}},\ \href {https://doi.org/10.1063/1.2336422} {10.1063/1.2336422} (\bibinfo {year} {2006})\BibitemShut {NoStop}%
\bibitem [{\citenamefont {Myra}\ \emph {et~al.}(2011)\citenamefont {Myra}, \citenamefont {Russell}, \citenamefont {D'Ippolito}, \citenamefont {Ahn}, \citenamefont {Maingi}, \citenamefont {Maqueda}, \citenamefont {Lundberg}, \citenamefont {Stotler}, \citenamefont {Zweben}, \citenamefont {Boedo},\ and\ \citenamefont {Umansky}}]{Myra2011}%
  \BibitemOpen
  \bibfield  {author} {\bibinfo {author} {\bibfnamefont {J.~R.}\ \bibnamefont {Myra}}, \bibinfo {author} {\bibfnamefont {D.~A.}\ \bibnamefont {Russell}}, \bibinfo {author} {\bibfnamefont {D.~A.}\ \bibnamefont {D'Ippolito}}, \bibinfo {author} {\bibfnamefont {J.~W.}\ \bibnamefont {Ahn}}, \bibinfo {author} {\bibfnamefont {R.}~\bibnamefont {Maingi}}, \bibinfo {author} {\bibfnamefont {R.~J.}\ \bibnamefont {Maqueda}}, \bibinfo {author} {\bibfnamefont {D.~P.}\ \bibnamefont {Lundberg}}, \bibinfo {author} {\bibfnamefont {D.~P.}\ \bibnamefont {Stotler}}, \bibinfo {author} {\bibfnamefont {S.~J.}\ \bibnamefont {Zweben}}, \bibinfo {author} {\bibfnamefont {J.}~\bibnamefont {Boedo}},\ and\ \bibinfo {author} {\bibfnamefont {M.}~\bibnamefont {Umansky}},\ }\bibfield  {title} {\bibinfo {title} {Reduced model simulations of the scrape-off-layer heat-flux width and comparison with experiment},\ }\bibfield  {journal} {\bibinfo  {journal} {Physics of Plasmas}\ }\textbf {\bibinfo {volume} {18}},\ \href
  {https://doi.org/10.1063/1.3526676} {10.1063/1.3526676} (\bibinfo {year} {2011})\BibitemShut {NoStop}%
\bibitem [{\citenamefont {Oldham}\ and\ \citenamefont {Spanier}(1974)}]{OldhamSpanier1974}%
  \BibitemOpen
  \bibfield  {author} {\bibinfo {author} {\bibfnamefont {K.~B.}\ \bibnamefont {Oldham}}\ and\ \bibinfo {author} {\bibfnamefont {J.}~\bibnamefont {Spanier}},\ }\href@noop {} {\emph {\bibinfo {title} {The Fractional Calculus: Theory and Applications of Differentiation and Integration to Arbitrary Order}}},\ \bibinfo {edition} {1st}\ ed.\ (\bibinfo  {publisher} {Academic Press},\ \bibinfo {year} {1974})\BibitemShut {NoStop}%
\bibitem [{\citenamefont {Ortigueira}\ and\ \citenamefont {Coito}(2004)}]{Ortigueira2004}%
  \BibitemOpen
  \bibfield  {author} {\bibinfo {author} {\bibfnamefont {M.~D.}\ \bibnamefont {Ortigueira}}\ and\ \bibinfo {author} {\bibfnamefont {F.}~\bibnamefont {Coito}},\ }\bibfield  {title} {\bibinfo {title} {From differences to derivatives},\ }\href@noop {} {\bibfield  {journal} {\bibinfo  {journal} {Fractional Calculus and Applied Analysis}\ }\textbf {\bibinfo {volume} {7}} (\bibinfo {year} {2004})}\BibitemShut {NoStop}%
\bibitem [{\citenamefont {Podlubny}(1998)}]{Podlubny1998}%
  \BibitemOpen
  \bibfield  {author} {\bibinfo {author} {\bibfnamefont {I.}~\bibnamefont {Podlubny}},\ }\href@noop {} {\emph {\bibinfo {title} {Fractional Differential Equations: An Introduction to Fractional Derivatives, Fractional Differential Equations, to Methods of their Solution and some of their Applications}}},\ \bibinfo {edition} {1st}\ ed.\ (\bibinfo  {publisher} {Academic Press},\ \bibinfo {year} {1998})\BibitemShut {NoStop}%
\bibitem [{\citenamefont {Çelik}\ and\ \citenamefont {Duman}(2012)}]{Celik2012}%
  \BibitemOpen
  \bibfield  {author} {\bibinfo {author} {\bibfnamefont {C.}~\bibnamefont {Çelik}}\ and\ \bibinfo {author} {\bibfnamefont {M.}~\bibnamefont {Duman}},\ }\bibfield  {title} {\bibinfo {title} {Crank-nicolson method for the fractional diffusion equation with the riesz fractional derivative},\ }\href {https://doi.org/10.1016/j.jcp.2011.11.008} {\bibfield  {journal} {\bibinfo  {journal} {Journal of Computational Physics}\ }\textbf {\bibinfo {volume} {231}},\ \bibinfo {pages} {1743} (\bibinfo {year} {2012})}\BibitemShut {NoStop}%
\end{thebibliography}%
\clearpage
\end{document}